\documentclass[amsfonts,amssymb,twocolumn,showpacs]{revtex4}
\usepackage{mathrsfs}
\usepackage{graphicx}

\begin{document}

\title{Nonlinear quantum interferometry with Bose condensed atoms}

\author{Chaohong Lee}
\altaffiliation{Electronic address: chleecn@gmail.com}

\author{Jiahao Huang}

\author{Haiming Deng}

\author{Hui Dai}

\author{Jun Xu}

\affiliation{State Key Laboratory of Optoelectronic Materials and Technologies,
School of Physics and Engineering, Sun Yat-Sen University, Guangzhou 510275,
China}

\date{\today}

\begin{abstract}
In quantum interferometry, it is vital to control and utilize nonlinear interactions for achieving high-precision measurements. Attribute to their long coherent time and high controllability, ultracold atoms including Bose condensed atoms have been widely used for implementing quantum interferometry. Here, we review the recent progresses in theoretical studies of quantum interferometry with Bose condensed atoms. In particular, we focus on the nonlinear phenomena induced by the atom-atom interaction and how to control and utilize these nonlinear phenomena. Under the mean-field description, due to the atom-atom interaction, matter-wave solitons appear in the interference patterns, and macroscopic quantum self-trapping exists in the Bose-Josephson junctions. Under the many-body description, the atom-atom interaction can generate non-classical entanglement, which may be utilized to achieve high-precision measurements beyond the standard quantum limit.

\vspace{2mm}

\end{abstract}

\maketitle

\tableofcontents

\section{Introduction}

\noindent It is well known that an ambitious goal of quantum physics is to control and exploit quantum coherence and entanglement.  Linear superposition of multiple quantum states is a perfect description for quantum coherence. Macroscopic quantum coherence (MQC), the quantum coherence among an ensemble of particles, may lead to novel phenomena never existed in single-particle systems and new applications in quantum interferometry. The systems of Bose-Einstein condensates (BECs)~\cite{Griffin-Book-1995,Pethick-Book-2002,Pitaevskii-Book-2003,Ueda-Book-2010}, such as, superfluids, superconductors and Bose-Einstein condensed atoms, are typical matters of MQC.

Under ultralow temperature close to absolute zero, as the thermal motions are almost frozen, the quantum features become significant and robust. In the last 20 years, there appear dramatic progresses in our ability to control and manipulate atoms. Now, atomic quantum gases and atomic BECs can be easily prepared by the well-developed techniques of trapping and cooling~\cite{Cornell-RMP-2002,Ketterle-RMP-2002}, and their degrees of freedom can also be accurately manipulated by applying laser and magnetic fields. In addition to new perspectives for the study of many-body quantum physics~\cite{Lewenstein-AdvPhys-2007,Bloch-RMP-2008}, these progresses offer new possibilities of atomic interferometers of unprecedented sensitivity~\cite{Cronin-RMP-2009}. In particular, nonlinear dynamics, quantum entanglement and quantum metrology meet together in systems of Bose condensed atoms~\cite{Sorensen-Nature-2001,Lee-PRL-2006,Boixo-PRL-2008,Pezze-PRL-2009, Gross-Nature-2010,Riedel-Nature-2010}, in which nonlinear dynamics may generate quantum entanglement and quantum entanglement may be used to implement high-precision quantum metrology.

This article is a review on recent progresses in theoretical studies in nonlinear quantum interferometry with Bose-Einstein condensed atoms. This section is an introduction for some fundamental conceptions for quantum interferometry, atomic BECs and nonlinear dynamics in BECs. In the next section, we present a description for MQC of Bose condensed atoms and discuss how to implement quantum interferometry via atomic matter-waves, in particular, the Bose-Josephson junctions. In the third section, we show how to realize many-body quantum interferometry via an ensemble of condensed atoms. The last section summarizes this article and discusses some perspectives in this field.

\subsection{Quantum interferometry}

\noindent In an optical interferometer, two or more light beams are coherently combined for interference and then one can extract their relative phase from their interference patterns. Optical interferometers have been extensively used for precision measurements, surface diagnostics, astrophysics, seismology, etc. There are several configurations for optical interferometers, for an example, an optical Mach-Zehnder interferometer splits a laser beam into two beams propagating along two paths and then recombines the two beams for interference and extracting their relative phase.

Beyond the conventional interferometry via classical waves, the quantum interferometry uses the wave nature of particles to achieve higher precision limit which can not be reached by classical interferometry. For an example, in a quantum Mach-Zehnder interferometer, a quantum state is transferred into a coherent superposition of two states by an unitary operation, then the two states accumulate different phases in a free time evolution, and lastly the two states are coherently recombined for interference and extracting their relative phase. The quantum interferometry has demonstrated by various systems from individual microscopic particles (photons, neutrons, electrons, atoms, molecules, etc) to macroscopic objects of an ensemble of microscopic particles (superfluids, superconductors, atomic BECs, etc).

It has demonstrated that quantum states in particular entangled states offer unprecedented advantages for metrology~\cite{Giovannetti-Science-2004,Ye-Science-2008,Appel-PNAS-2009, Giovannetti-Nphoton-2011,Escher-NatPhys-2011,Ma-arXiv-2011, Benatti-AnnPhys-2010, Benatti-JPB-2011, Argentieri-arXiv-2011, Zwierz-PRL-2010}. The quantum states have been extensively used in high-precision measurements, such as, quantum states of photons have been used for high-resolution imaging, quantum states of spins have been used for high-resolution measurements of magnetic field, and quantum states of atoms have been used to build clocks of ultimate accuracy. Moreover, multi-atom systems in particular entangled multi-atom systems, such as ultracold atoms in optical lattices and quantum atomic gases, provide new opportunity for designing next-generation atomic clocks~\cite{Derevianko-RMP-2011}.

\subsection{Atomic Bose-Einstein condensates and nonlinear dynamics}

\noindent In an atomic BEC, the matter-waves of individual atoms overlap each other and therefore all atoms lose their distinguishability. The indistinguishability of individual atoms means the appearance of MQC. Under such a low temperature, the atom-atom interaction in a BEC is dominated by the s-wave scattering. Under the mean-field description~\cite{Griffin-Book-1995,Pethick-Book-2002,Pitaevskii-Book-2003,Ueda-Book-2010}, an atomic BEC obeys a nonlinear Schr\"{o}dinger equation, the Gross-Pitaevskii equation, in which the nonlinear strength is proportional to the s-wave scattering length. Due to the intrinsic nonlinearity from the atom-atom interaction, the de Broglie matter waves of Bose condensed atoms have led to the development of nonlinear and quantum optics with atoms~\cite{Rolston-Nature-2002}, which are analogues of conventional nonlinear and quantum optics with photons.

It has demonstrated the existence of many nonlinear phenomena such as four-wave mixing, solitons and chaos in atomic BECs.  In an optical four-wave mixing, a fourth light wave is produced by mixing three different light waves under the condition of energy and momentum conservation. Similarly, four-wave mixing process may take place in atomic BECs~\cite{Deng-Nature-1999}, in which three different matter waves of atoms mix to produce a fourth matter wave of atoms. Solitons are stable solutions which maintain their shape unchanged during their propagation. Due to the balance between dispersion and nonlinear atom-atom interaction, matter wave solitons may appear in atomic BECs. The dark~\cite{Denschlag-Science-2000,Burger-PRL-1999} and bright solitons~\cite{Khaykovich-Science-2002,Strecker-Nature-2002} have been observed in atomic BECs with repulsive and attractive atom-atom interactions, respectively. Due to the intrinsic nonlinearity from the mean-field interaction, macroscopic quantum chaos has been theoretically predicted in time driven systems~\cite{Lee-PRA-2001,Hai-PRE-2002} and high-dimensional systems of atomic BECs~\cite{Thommen-PRL-2003,Summy-PhysicaScripta-2004}. The strength of nonlinearity (i.e. the s-wave scattering length) can be tuned by Feshbach resonances~\cite{Chin-RMP-2010} and so that the nonlinear dynamics can be modulated~\cite{Pollack-PRL-2009, Vidanovic-PRA-2011}.

\subsection{Quantum interferometry via Bose condensed atoms}

\noindent Since the first realization of Bose-Einstein condensation in ultracold atomic gases, the condensed atoms have been extensively used for performing quantum interferometry~\cite{Cronin-RMP-2009,Kasevich-BookChapter-2001,Minardi-BookChapter-2001,Kasevich-Science-2002}. There are two fundamental schemes for quantum interferometry with condensed atoms: space-domain interferometry and time-domain interferometry. In the space-domain scheme, two or more atomic BECs at different spatial positions are released for expansion and then the matter waves from different condensates overlap each other. In the time-domain scheme, by means of a double separated oscillator technique, the Ramsey oscillations of condensed atoms involving two or more internal hyperfine levels have been demonstrated.

\section{Macroscopic quantum coherence and atomic matter-wave interferometry}

\subsection{Macroscopic quantum coherence and mean-field description}

\noindent It is well known that wave-like properties of single particles such as photons, electrons and atoms have been demonstrated in various experiments. Fantastically, for an atomic BEC, the collection of atoms may also behave as an entire object of wave-like properties~\cite{Griffin-Book-1995,Pethick-Book-2002,Pitaevskii-Book-2003,Ueda-Book-2010}. These wave-like properties among an ensemble of particles are a signature of macroscopic quantum coherence (MQC), which relates to the superposition of multiple macroscopic quantum states.

In the quantum field theory, an ensemble of ultracold Bose atoms confined within an external potential obeys the many-body Hamiltonian~\cite{Griffin-Book-1995,Pethick-Book-2002,Pitaevskii-Book-2003,Ueda-Book-2010,Giorgini-RMP-1999,Leggett-RMP-2001},
\begin{eqnarray}
\hat{H}&=&\int d \mathbf{r}\hat{\Psi}^{\dag}(\mathbf{r})\left[-\frac{\hbar^{2}\nabla^{2}}{2m} +V_{ext}(\mathbf{r})\right]\hat{\Psi}(\mathbf{r})\nonumber\\
&+&\frac{1}{2}\int d \mathbf{r} d \mathbf{r}'\hat{\Psi}^{\dag}(\mathbf{r})\hat{\Psi}^{\dag}(\mathbf{r}') V(\mathbf{r}-\mathbf{r}')\hat{\Psi}(\mathbf{r}')\hat{\Psi}(\mathbf{r}),
\label{pythag}
\end{eqnarray}
with $\hat{\Psi}^{\dag}$ and $\hat{\Psi}$ denoting bosonic fields of atoms. Here, $m$ is the atomic mass, $V_{ext}$ stands for the external potential, and $V(\mathbf{r}-\mathbf{r}')$ describes the interaction between two atoms at positions $\mathbf{r}$ and $\mathbf{r}'$.

The time evolution of the field operators is given by the Schr\"{o}dinger equation,
\begin{eqnarray}
&i\hbar&\frac{\partial }{\partial t}\hat{\Psi}(\mathbf{r},t)= \left[-\frac{\hbar^{2}\nabla^{2}}{2m}+V_{ext}(\mathbf{r})\right.\nonumber\\
&~~~&\left.+\int d \mathbf{r}'\hat{\Psi}^{\dag}(\mathbf{r}',t) V(\mathbf{r}-\mathbf{r}') \hat{\Psi}(\mathbf{r}',t)\right]\hat{\Psi}(\mathbf{r},t).
\label{pythag}
\end{eqnarray}
Solving the above equation for an ensemble of particles involves heavy numerical work. For an atomic BEC dominated by condensed atoms, mean-field (MF) theory allows one to well understand most behaviors related to MQC. However, as will be discussed in Section 3, the theoretical treatment must go beyond MF and take into account quantum fluctuations if non-classical states such as squeezed states and macroscopic coherent superpositions are encountered.

The field operator $\hat{\Psi}(\mathbf{r},t)$  can be expressed in form of
\begin{equation}
 \hat{\Psi}(\mathbf{r},t)={\Phi}(\mathbf{r},t)+\hat{\Psi}'(\mathbf{r},t),
\label{pythag}
\end{equation}
where ${\Phi}(\mathbf{r},t)\equiv \langle \hat{\Psi}(\mathbf{r},t)\rangle$ is called as the order parameter or ¡°wave-function of the condensate" and $\hat{\Psi'}(\mathbf{r},t)$ describes the fluctuations around the condensate state. Therefore, the condensate density reads as $n_{0}(\mathbf{r},t) = \left|{\Phi}(\mathbf{r},t)\right|^{2}$.

Under ultra-low temperatures, if the average inter-particle spacing is sufficiently large, the atom-atom interaction is dominated by the s-wave scattering. That is, the atom-atom interaction is effectively described by a ``contact interaction",
\begin{equation}
V(\mathbf{r}'-\mathbf{r}) = g \delta(\mathbf{r}'-\mathbf{r}),
\label{pythag}
\end{equation}
with $g \equiv \frac{4\pi\hbar^{2}a}{m}$ and $a$ being the s-wave scattering length. Physically, $g>0$ and $g<0$ correspond to repulsive and attractive interactions, respectively. Inserting the delta interaction potential into the many-body quantum Hamiltonian (1), one can find
\begin{eqnarray}
\hat{H}&=&\int d \mathbf{r}\hat{\Psi}^{\dag}(\mathbf{r})\left[-\frac{\hbar^{2}\nabla^{2}}{2m} +V_{ext}(\mathbf{r})\right]\hat{\Psi}(\mathbf{r})\nonumber\\
&&+\frac{g}{2}\int d \mathbf{r} \hat{\Psi}^{\dag}(\mathbf{r})\hat{\Psi}^{\dag}(\mathbf{r}) \hat{\Psi}(\mathbf{r})\hat{\Psi}(\mathbf{r}).
\label{pythag}
\end{eqnarray}
For an atomic gas dominated by condensed atoms, the fluctuation part $\hat{\Psi'}(\mathbf{r},t)$ is negligible.

Under conditions of ultralow temperature and dilute density of atoms, that is, (a) $\hat{\Psi}(\mathbf{r},t)\approx {\Phi}(\mathbf{r},t)$ of very small fluctuations, (b) the inter-atom interaction is a contact interaction, and (c) the average inter-particle distance is much larger than the s-wave scattering length, an atomic BEC can be described by a mean-field Hamiltonian,
\begin{eqnarray}
H_{MF}&=&\int d \mathbf{r}\Phi^{*}(\mathbf{r})\left[-\frac{\hbar^{2}\nabla^{2}}{2m} +V_{ext}(\mathbf{r})\right]\Phi(\mathbf{r})\nonumber\\
&&+\frac{g}{2}\int d \mathbf{r} {\Phi}^{*}(\mathbf{r}){\Phi}^{*}(\mathbf{r}) \Phi(\mathbf{r})\Phi(\mathbf{r}).
\end{eqnarray}
The time evolution of the condensate wave-function $\Phi(\mathbf{r})$ obeys a nonlinear Schr\"{o}dinger equation, the Gross-Pitaevskii equation (GPE),
\begin{equation}
i\hbar\frac{\partial }{\partial t}\Phi (\mathbf{r},t)=\left[-\frac{\hbar^{2}\nabla^{2}}{2m}+V_{ext}(\mathbf{r}) +g|{\Phi}(\mathbf{r},t)|^{2}\right]\Phi(\mathbf{r},t).
\label{pythag}
\end{equation}
Here, the term of $g$ describes the nonlinear mean-field interaction between atoms. For a stationary state, $\Phi(\mathbf{r},t)=\phi(\mathbf{r})\exp(-i\mu t/\hbar)$, the spatial wave function obeys a time-independent GPE,
\begin{equation}
\mu\phi(\mathbf{r})=\left[-\frac{\hbar^{2}\nabla^{2}}{2m}+V_{ext}(\mathbf{r}) +g{\phi}^{2}(\mathbf{r})\right]\phi£¨\mathbf{r}),
\label{pythag}
\end{equation}
with $\mu$ denoting the chemical potential.

\subsection{Atomic matter-wave interference}

\noindent The first interference experiment of two atomic BECs was implemented in 1997 by Ketterle's group at MIT~\cite{Andrews-Science-1997}. In this experiment, two atomic BECs released from a double-well potential freely expand and then form clear interference patterns. Generally speaking, the double-well potential can be created by superposing a potential barrier on a harmonic potential. In MIT's experiment, the harmonic trap is a magnetic potential and the barrier is provided by a blue-detuned far-off-resonance laser beam. There are several other schemes for forming double-well potentials, such as, radio-frequency-induced adiabatic potentials created by atom chips~\cite{Shin-PRL-2004,Schumm-NatPhys-2005,Schumm-QIP-2006} and superposing a weakly optical lattices on a strongly harmonic trap~\cite{Albiez-PRL-2005,Gati-JPB-2007}.

There are two different methods for preparing two BECs in a double-well potential. The first method is loading cold atoms into a double-well potential and obtaining two BECs via evaporative cooling. The second method is preparing a BEC in a harmonic trap and then splitting it into two BECs via increasing the potential barrier.

Interference patterns will appear if atomic matter waves from different condensates may reach same spatial positions. A natural way is switching off the external potential. The experimental data shows the interference patterns differ from the ones of two point-like monochromatic sources and two point-like pulsed sources. This indicates that the two BECs released from a double-well potential can not be simply treated as two point-like monochromatic sources or two point-like pulsed sources. In general, for particles from two point-like sources, their fringe period of the interference pattern is the de Broglie wavelength associated with the relative motion of particles, $\lambda =h/p = ht/(md)$, where $h$ is Planck's constant, $p$ is the relative momentum, $t$ is the time, $d$ is the distance between two sources and $m$ is the particle mass. For the case of two BECs released from a double-well potential, the fringe period formulae for two point-like sources is only valid for the central fringe~\cite{Andrews-Science-1997}.

For simplicity, we consider an elongated system of condensed atoms. That is, the atoms are confined by a strong transverse confinement, $m\omega^{2}_{\rho} (y^{2}+z^{2})/2$. Integrating the transverse coordinates, the elongated system can be described by a one-dimensional (1D) GPE~\cite{Perez-Garcia-PRA-1998,Carr-PRL-2004}:
\begin{equation}
i \hbar \frac{\partial}{\partial t} \Phi (x,t) = H_{0} \Phi (x,t)
+ \lambda \left|\Phi (x,t) \right|^{2} \Phi (x,t),
\end{equation}
where $H_0 = -(\hbar^{2}/2m)(\partial^{2}/\partial x^{2}) +
V(x,t)$, $\lambda$ characterizes the nonlinear MF interaction, and
$V(x,t)$ is an external potential.

Usually, the condensate wave-function $\Phi(\mathbf{r},t)$ is normalized to the total number of condensed atoms $N$. To compare with the conventional Schr\"{o}dinger equation of no MF interaction, we normalize $\Phi(x,t)$ for our 1D model to one and its nonlinearity strength $\lambda = 2N a_{s} \omega_{\rho} \hbar$ is determined by $N$, the s-wave scattering length $a_{s}$ and the transverse trapping frequency $\omega_{\rho}$~\cite{Carr-PRL-2004}. For convenience, we introduce the dimensionless model equation by choosing the natural units of $m = \hbar =1$. The external potential $V(x,t)$ is formed by a superposition of a time-independent harmonic potential and a time-dependent Gaussian barrier, see Fig. 1. That is,
\begin{equation}
 V(x,t) = \frac{1}{2} \omega^{2} x^{2} + B(t) \cdot \exp \left(-\frac{x^{2}}{2d^{2}}\right),
\end{equation}
where $\omega$ is the trapping frequency, $d$ is the barrier width, and the barrier height $B(t)$ is a function of time. If $B(t)>\omega^2d^2$, $V(x,t)$ is a double-well potential with two minima at $x=\pm d\sqrt{2\ln[B/(d^2\omega^2)]}$. In the splitting process, the barrier height is adiabatically increased and therefore the single condensate in a harmonic trap splits into two condensates in a deep double-well potential. To observe the interference pattern, one has to switch off the potential barrier at least and let the two condensates overlap each other.

\begin{figure}[htp]
\center
\includegraphics[width=1.0\columnwidth]{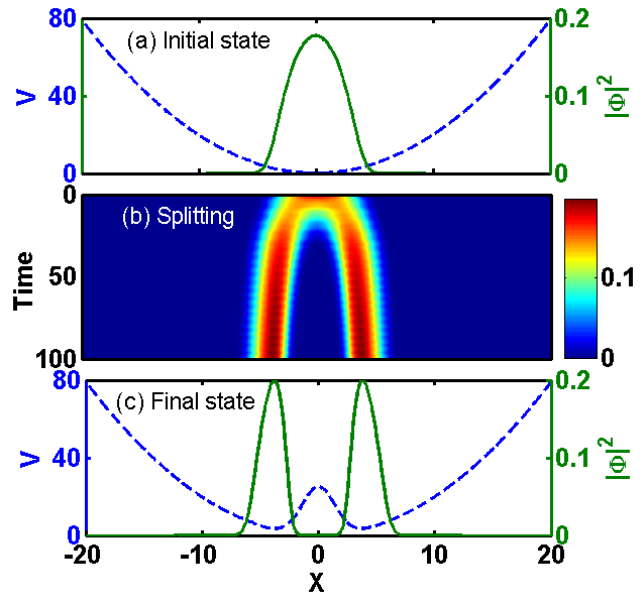}
\caption{The splitting process from a single condensate to two condensates. (a) The initial state (green solid curve) at time $t=0$ in a harmonic trap (blue dashed curve); (b) the density evolution $\left|\Phi(x,t)\right|^2$ for the splitting process; and (c) the final state (green solid curve) at time $t=100$ in a double-well potential (blue dashed curve).}
\end{figure}

In Fig. 1, we show our numerical simulation for the splitting process. Initially, the condensate stays in the ground state for the harmonic trap. Then the system is transferred into the ground state for a deep double-well potential by slowly increasing the barrier height. If the barrier is sufficiently high, the condensates in two wells almost have no overlap and so that they can be looked as two independent condensates. In our calculation, we choose the interaction strength $\lambda=20$, the barrier width $d=\sqrt{2}$, the trap frequency $\omega=0.2\pi$, and the barrier height $B$ is linearly increased from zero to 25.

\begin{figure}[htp]
\center
\includegraphics[width=1.0\columnwidth]{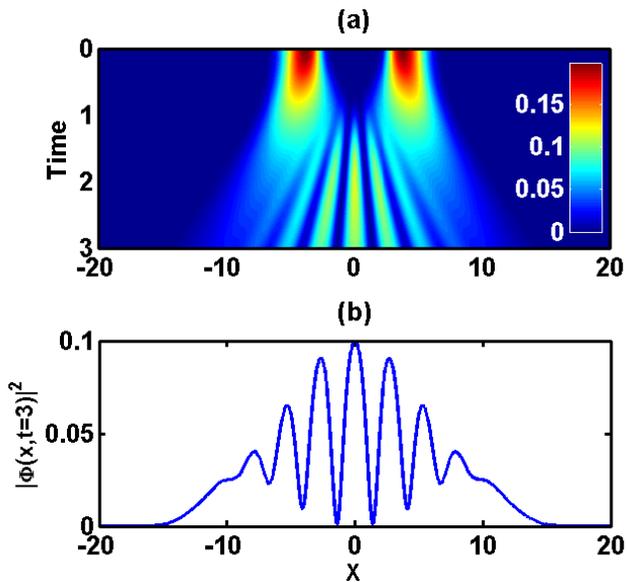}
\caption{Interference fringes of two freely expanding condensates. (a) The density evolution $\left|\Phi(x,t)\right|^2$ after switching off the external potential $V(x,t)$; and (b) the interference fringe at time $t=3$.}
\end{figure}

The external potential $V(x,t)$ is suddenly switched off when two condensates have been prepared in the double-potential. Thus the two condensates freely expand and interference fringes gradually appear when atoms from different condensates reach the same position. In Fig. 2, we show our numerical simulation for how the interference fringes appear. In our calculation, we choose the final state of the splitting process shown in Fig. 1 as the initial state before switching off the external potential.

\subsection{Nonlinear excitations in atomic matter-wave interference}

\noindent The MF theory for atomic BECs predicts that solitons may exist in 1D BEC systems. If the s-wave scattering length is negative, the atom-atom interaction is attractive and acts as the role of self-focusing in nonlinear optics, solitons appear as shape-maintaining wavepackets in propagation. If the s-wave scattering length is positive, the atom-atom interaction is repulsive and acts as the role of self-defocusing in nonlinear optics, solitons appear as shape-maintaining notches in propagation.

In matter-wave interference experiments of atomic BECs, due to the intrinsic atom-atom interaction, the interference patterns differ from the ones of linear waves and nonlinear excitations may gradually appear. Theoretically, it has demonstrated that dark solitons can be generated by collision~\cite{Reinhardt-JPB-1997,Negretti-JPB-2004} and quantum tunneling~\cite{Lee-JPB-2007} in double-well BEC interferometers. Additionally, nonlinear effects in the BEC interference are studied by using exact solutions of the one-dimensional GPE~\cite{Liu-PRL-2000}. Experimentally, for atomic BECs with repulsive MF interaction, dark solitons are gradually generated by merging and splitting BECs~\cite{Jo-PRL-2007,Weller-PRL-2008,Chang-PRL-2008,Shomroni-NatPhys-2009}.

The soliton generation depends on the initial phase difference and how fast the recombination occurs. In a sufficiently slow recombination, if the initial state is the ground state for the double-well potential of no phase difference, the condensate is adiabatically transferred into its ground state for the final harmonic trap. However, if the initial state is the first-excited state for the double-well potential of $\pi$ phase difference, the condensate will be adiabatically transferred into its first-excited state for the final harmonic trap. The notch of this excited state can be looked as a dark soliton. Usually, more number of solitons will be generated in a faster recombination~\cite{Lee-JPB-2007}.

\begin{figure}[htp]
\center
\includegraphics[width=1.0\columnwidth]{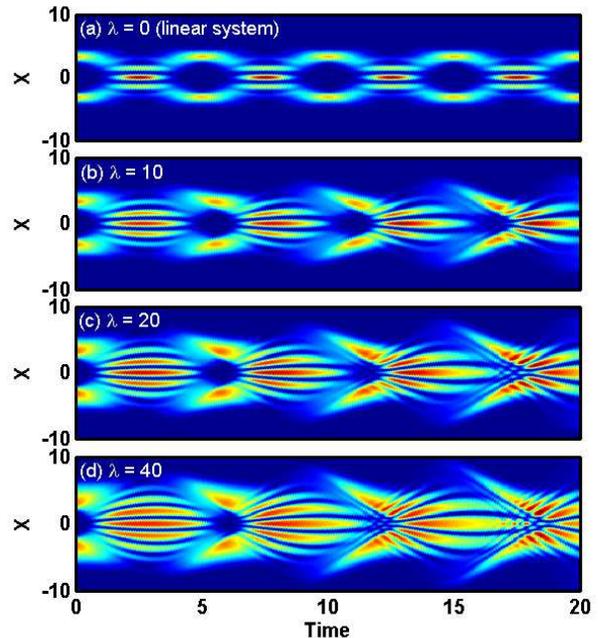}
\caption{In-trap interference and nonlinear excitations. The density evolutions $\left|\Phi(x,t)\right|^2$ for $\lambda = 0$ (linear system), $\lambda = 10$, $\lambda = 20$ and  $\lambda = 40$ are shown in rows (a), (b), (c) and (d), respectively. All initial states are chosen as the ground states for BECs within a double-well potential of parameters $\omega=0.2\pi$, $d = 1$ and $B = 40$.}
\end{figure}

Based upon the 1D GPE (7), we explore how dark solitons appear via numerical integration, see Fig. 3. The initial states are set as the condensate ground states for the external potential $V(x,t=0)$ with parameters $\omega=0.2\pi$, $d =1$ and $B\left(t=0\right) = 40$. Following the experiments on recombination from a double-well potential into a harmonic potential~\cite{Jo-PRL-2007,Weller-PRL-2008}, we only switch off the potential barrier and keep the trap frequency $\omega$ unchanged. There are several schemes for switching off the potential barrier. In our simulation, similar to the matter-wave interference in free space, we suddenly switch off the potential barrier, that is, $B\left(t>0\right) = 0$). Due to the harmonic forces, two condensates collide again and again. The interference patterns strongly depend on the interaction strength. For a linear system ($\lambda = 0$) of no inter-particle interaction, the interference patterns periodically oscillate with the period for the harmonic trap. For the nonlinear systems of repulsive interactions ($\lambda > 0$), the periodicity for the interference patterns is destroyed by the nonlinear inter-particle interactions and dark solitons are gradually generated. In general, solitons gradually appear from the overlapped regions and more solitons are generated in more strongly nonlinear systems.

\subsection{Bose-Josephson junction (BJJ)}

\noindent In 1962, Brian Josephson predicted the existence of supercurrent through an insulator barrier between two superconductors~\cite{Josephson-PLA-1962,Josephson-RMP-1974}. This tunnelling effect of electrons is named as the Josephson effect. The device of two superconductors linked by an insulator barrier is known as a superconductor Josephson junction (SJJ), which has been extensively used to build quantum interference devices for sensitive measurements.

To detect and exploit the MQC among BECs of atoms, similar to the SJJs, it is natural to link different condensates with Josephson couplings. The peculiar tunnelling phenomena in Josephson coupled BECs are called as Bose-Josephson effects, and the physical systems of two BECs linked by Josephson couplings are called as Bose-Josephson junctions (BJJs). There are two different types of BJJs: the external and internal BJJs~\cite{Leggett-RMP-2001}, see their schematic diagrams in Fig. 4. An external BJJ~\cite{Shin-PRL-2004, Schumm-NatPhys-2005, Schumm-QIP-2006, Albiez-PRL-2005, Gati-JPB-2007} involves two condensates in a double-well potential and the Josephson coupling is provided by the quantum tunnelling through the barrier between two wells, see panel (a) of Fig. 4. An internal BJJ~\cite{Hall-PRL-1998, Smerzi-EPJB-2003, Egorov-PRA-2011} involves a two-component condensate of atoms occupying two hyperfine levels which are coupled by the external fields, see panel (b) of Fig. 4. Besides the Josephson effects in two-mode systems, non-Abelian Josephson effects have been predicted in a spin-2 BEC of five modes~\cite{Qi-PRL-2009}. Furthermore, the Josephson oscillations have been demonstrated in an array of BECs in optical lattices~\cite{Cataliotti-Science-1998,Anderson-Science-1998,Morsch-RMP-2006}.
\vspace{3mm}
\begin{figure}[htp]
\center
\includegraphics[width=1.0\columnwidth]{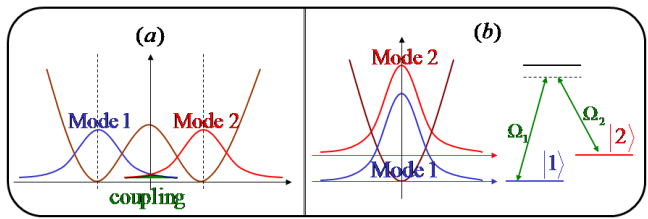}
\caption{Schematic diagrams for Bose-Josephson junctions: (a) an external Bose-Josephson junction linked by quantum tunnelling, and (b) an internal Bose-Josephson junction via a two-component BEC linked by Raman fields.}
\end{figure}

Because of the nonlinear MF interaction, several exotic macroscopic quantum phenomena have been demonstrated, such as macroscopic quantum self-trapping (MQST)~\cite{Kenkre-PRB-1986,Smerzi-PRL-1997,Raghavan-PRA-1999,Kuang-PRA-2000,Hu-PRA-2000} and macroscopic quantum chaos~\cite{Lee-PRA-2001,Hai-PRE-2002}. Moreover, in contrast to the linear Landau-Zener (LZ) tunneling in two-state single-particle quantum systems, loop structures may appear in the nonlinear LZ tunneling in two-mode BEC systems~\cite{Wu-PRA-2000,Liu-PRA-2002,Jona-Lasinio-PRL-2003}. Actually, these loop structures and MQST are two different sides of the bistability induced by the nonlinear MF interaction.

In this subsection, under the MF description, we will focus on discussing the macroscopic quantum phenomena in BJJs. At first, we introduce an unified model for BJJs. Then, we discuss the dynamics in time-independent systems, such as, Rabi oscillation, MQST and static bifurcation. At last, we analyze the dynamics in time-dependent systems, such as, chaos and slow passages across the critical point.

\subsubsection{Two-mode approximation}

\noindent In the MF theory, BJJs can be treated as multiple coupled modes of macroscopic matter waves~\cite{Lee-JPB-2007,Ostrovskaya-PRA-2000}. For an external BJJ with a sufficiently high potential barrier, the condensate wave-function $\Phi(r,t)$ can expanded by two Wannier states for the two wells~\cite{Smerzi-PRL-1997,Milburn-PRA-1997}. That is,
\begin{equation}\label{dhm2}
  \Phi(\mathbf{r},t)=\psi_1(t)\phi_1(\mathbf{r})+ \psi_2(t)\phi_2(\mathbf{r}),
\end{equation}
where the spatial functions $\phi_j(\mathbf{r})$ (j = 1 and 2) are the two Wannier states and the time-dependent amplitudes $\psi_j(t)$ are complex numbers. As the condensate wave-function $\Phi(r,t)$ is normalized to the total number of condensed atoms $N$ and each Wannier state is normalized to one, we have $\left|\psi_1(t)\right|^2 + \left|\psi_2(t)\right|^2 = N$.

Under the two-mode approximation (11), integrating all spatial coordinates, the MF Hamiltonian read as,
\begin{eqnarray}
 H_{MF} = && - J \left(\psi_1^*\psi_2+\psi_2^*\psi_1\right) + \varepsilon_1\left|\psi_1\right|^2 +\varepsilon_2\left|\psi_2\right|^2\nonumber\\
 &&+ \frac{1}{2}U_{11}\left|\psi_1\right|^4+\frac{1}{2}U_{22}\left|\psi_2\right|^4\nonumber\\
 &&+ 2 U_{12}\left|\psi_1\right|^2\left|\psi_2\right|^2\nonumber\\
 &&+ \frac{1}{2}U_{11 \Leftrightarrow 22} \left[\left(\psi_1^*\psi_2\right)^2 + \left(\psi_2^*\psi_1\right)^2\right]\nonumber\\
 &&+ U_{11 \Leftrightarrow 12}\left|\psi_1\right|^2 \left(\psi_1^*\psi_2+\psi_2^*\psi_1\right)\nonumber\\
 &&+ U_{22 \Leftrightarrow 12}\left|\psi_2\right|^2 \left(\psi_1^*\psi_2+\psi_2^*\psi_1\right),
\end{eqnarray}
with the inter-mode tunnelling strength,
$$ J = -\int d \mathbf{r} \left[\phi_2^*(\mathbf{r})
(-\frac{\hbar^{2}\nabla^{2}}{2m}+V)\phi_1(\mathbf{r})\right],$$
the zero-point energy for atoms in the j-th mode,
$$
\varepsilon_j = \int d\mathbf{r} \left[\phi_j^*(\mathbf{r}) (-\frac{\hbar^{2}\nabla^{2}}{2m}+V) \phi_j(\mathbf{r}) \right],
$$
the intra-mode interaction strength,
$$U_{jj} = g\int d\mathbf{r} \left[\left|\phi_j(\mathbf{r})\right|^4\right],$$
the inter-mode interaction strength,
$$U_{12} = g\int d\mathbf{r} \left[\left|\phi_1(\mathbf{r})\right|^2 \left|\phi_2(\mathbf{r})\right|^2 \right],$$
and the inter-mode exchange collision interaction strength,
$$U_{jk \Leftrightarrow lm} = g\int d\mathbf{r} \left[\phi_j^*(\mathbf{r})\phi_k^*(\mathbf{r}) \phi_l(\mathbf{r})\phi_m(\mathbf{r}) \right].$$
In the exchange collision denoted by $jk \Leftrightarrow lm$, an atom from the j-th mode collides with an atom from the k-th mode and then these two atoms are transferred into an atom in the l-th mode and an atom in the m-th mode. For a sufficiently deep double-well potential, two Wannier states are well localized at the two well-centers, so that the high-order overlaps between two Wannier states are very small. This indicates that the terms in last four rows of Hamiltonian (12) can be ignored, that is,
\begin{eqnarray}
 H_{MF} = && - J \left(\psi_1^*\psi_2+\psi_2^*\psi_1\right) + \varepsilon_1\left|\psi_1\right|^2 +\varepsilon_2\left|\psi_2\right|^2\nonumber\\
 &&+ \frac{1}{2}U_{11}\left|\psi_1\right|^4+\frac{1}{2}U_{22}\left|\psi_2\right|^4.
\end{eqnarray}

For an internal BJJ, in which a two-component BEC involves atoms in two-hyperfine levels coupled by laser or radio-frequency fields, its Hamiltonian in quantum field theory reads as
\begin{equation}
\hat{H} = \hat{H}_{1}+\hat{H}_{2}+\hat{H}_{12},
\end{equation}
with the single component Hamiltonian for the first component,
\begin{eqnarray}
\hat{H}_{1}=&&\int\hat{\Psi}_{1}^{\dag}(\mathbf{r},t)(-\frac{\hbar^{2}\nabla^{2}}{2m}
       +V(\mathbf{r},t))\hat{\Psi}_{1}(\mathbf{r},t)d\mathbf{r}\nonumber\\
     &&+\frac{g_{11}}{2}\int\hat{\Psi}_{1}^{\dag}(\mathbf{r},t)\hat{\Psi}_{1}^{\dag}(\mathbf{r},t)
       \hat{\Psi}_{1}(\mathbf{r},t)\hat{\Psi}_{1}(\mathbf{r},t)d\mathbf{r}\nonumber\\
     &&- \frac{\Delta}{2} \int\hat{\Psi}_{1}^{\dag}(\mathbf{r},t)\hat{\Psi}_{1}(\mathbf{r},t)d\mathbf{r},\nonumber
\end{eqnarray}

the single component Hamiltonian for the second component,
\begin{eqnarray}
\hat{H}_{2}=&&\int\hat{\Psi}_{2}^{\dag}(\mathbf{r},t)(-\frac{\hbar^{2}\nabla^{2}}{2m}
       +V(\mathbf{r},t))\hat{\Psi}_{2}(\mathbf{r},t)d\mathbf{r}\nonumber\\
     &&+\frac{g_{22}}{2}\int\hat{\Psi}_{2}^{\dag}(\mathbf{r},t)
       \hat{\Psi}_{2}^{\dag}(\mathbf{r},t)\hat{\Psi}_{2}(\mathbf{r},t) \hat{\Psi}_{2}(\mathbf{r},t)d\mathbf{r}\nonumber\\
     &&+\frac{\Delta}{2}\int\hat{\Psi}_{2}^{\dag}(\mathbf{r},t) \hat{\Psi}_{2}(\mathbf{r},t)d\mathbf{r},\nonumber
\end{eqnarray}
and the inter-component interaction and the linear coupling between two components,
\begin{eqnarray}
\hat{H}_{12}=&&g_{12}\int\hat{\Psi}_{2}^{\dag}(\mathbf{r},t)\hat{\Psi}_{1}^{\dag}(\mathbf{r},t)
        \hat{\Psi}_{1}(\mathbf{r},t)\hat{\Psi}_{2}(\mathbf{r},t)d\mathbf{r}\nonumber\\
       &&-\frac{\hbar\Omega}{2}\int(\hat{\Psi}_{1}^{\dag}(\mathbf{r},t)\hat{\Psi}_{2}(\mathbf{r},t)
        +\hat{\Psi}_{2}^{\dag}(\mathbf{r},t)\hat{\Psi}_{1}(\mathbf{r},t))d\mathbf{r}.\nonumber
\end{eqnarray}
Here, $m$ is the single-atom mass, $g_{jk}=4\pi\hbar^{2}a_{jk}/m$ with $a_{jk}$ denoting the s-wave scattering length between atoms in hyperfine states $|j\rangle$ and $|k\rangle$, $V(\mathbf{r},t)$ is the external harmonic potential, $\Delta$ is the detuning to the resonant transition between the two hyperfine levels and $\Omega$ is the Rabi frequency. The symbols $\hat{\Psi}_{j}^{\dag}(\mathbf{r},t)$ and $\hat{\Psi}_{j}(\mathbf{r},t)$ are Bose creation and annihilation operators for atoms in $|j\rangle$, respectively.

The system can be described the MF theory if the condensed atoms dominate the whole system, i.e., $\hat{\Psi}_{j}(\mathbf{r},t) \approx \Phi_{j}(\mathbf{r},t)$. Under a sufficiently strong confinement, only the ground state for the external potential will be occupied and so that the condensate wave-functions can be expressed as~\cite{Williams-PRA-1999,Lee-PRA-2004}
\begin{equation}
\Phi_{j}(\mathbf{r},t) = \psi_{j}(t) \phi(\mathbf{r}),
\end{equation}
with $\phi(\mathbf{r})$ describes the spatial distribution and the complex amplitudes satisfy the normalization condition, $\left|\psi_{1}(t)\right|^2+\left|\psi_{2}(t)\right|^2 = N$. Here, $N$ is the total number of atoms. Inserting the approximation ansatz (15) into the Hamiltonian (14) and integrating all spatial coordinates, one will find that the complex amplitudes obey the following MF Hamiltonian,
\begin{eqnarray}
 H_{MF} = && - J \left(\psi_1^*\psi_2+\psi_2^*\psi_1\right) + \varepsilon_1\left|\psi_1\right|^2 +\varepsilon_2\left|\psi_2\right|^2\nonumber\\
 &&+ \frac{1}{2}U_{11}\left|\psi_1\right|^4+\frac{1}{2}U_{22}\left|\psi_2\right|^4\nonumber\\
 &&+ U_{12}\left|\psi_1\right|^2\left|\psi_2\right|^2,
\end{eqnarray}
with the Josephson coupling strength,
$$
J=\frac{\hbar\Omega}{2} \int d \mathbf{r} \left[\phi^*(\mathbf{r}) \phi(\mathbf{r})\right] =\frac{\hbar\Omega}{2},
$$
the single-particle energy for the first-component,
$$
\varepsilon_1 = \int d\mathbf{r} \left[\phi^*(\mathbf{r}) (-\frac{\hbar^{2}\nabla^{2}}{2m}+V-\frac{\Delta}{2}) \phi(\mathbf{r}) \right],
$$
the single-particle energy for the second-component,
$$
\varepsilon_2 = \int d\mathbf{r} \left[\phi^*(\mathbf{r}) (-\frac{\hbar^{2}\nabla^{2}}{2m}+V+\frac{\Delta}{2}) \phi(\mathbf{r}) \right],
$$
the intra-component interaction strength,
$$U_{jj} = g_{jj}\int d\mathbf{r} \left[\left|\phi(\mathbf{r})\right|^4\right],$$
and the inter-component interaction strength,
$$U_{12} = g_{12}\int d\mathbf{r} \left[\left|\phi(\mathbf{r})\right|^4 \right].$$

The time evolution of BJJs obeys two coupled differential equations
\begin{equation}
i\hbar\frac{d \psi_j(t)}{dt}=\frac{\partial H_{MF}}{\partial \psi_j^*(t)},
\end{equation}
with $H_{MF}$ given by the Hamiltonian (13) or (16). Obviously, the time evolution conserves the total number of atoms, $N = \left|\psi_{1}(t)\right|^2+\left|\psi_{2}(t)\right|^2$. This means that $N$ commutates with the Hamiltonian. Thus, one can omit the terms $O(N)$ and $O(N^2)$ in Hamiltonians (13) and (16) and rewrite them into an unified form,
\begin{equation}
H = \frac{\delta}{2}\left(n_2-n_1\right) + \frac{E_c}{8}\left(n_2-n_1\right)^2 -J\left(\psi_1^*\psi_2+\psi_2^*\psi_1\right),
\end{equation}
with $n_j=\psi_j^*\psi_j=\left|\psi_j\right|^2$, $\delta = \varepsilon_2 -\varepsilon_1 + N\left(U_{22} - U_{11}\right)/4$, and $E_c = U_{11} + U_{22}$ for external BJJs or $E_c = U_{11} + U_{22} - 2U_{12}$ for internal ones. Correspondingly, the time evolution of BJJs obeys the following two coupled discrete GPEs,
\begin{equation}
i\hbar\frac{d \psi_1}{dt} = -\frac{\delta}{2}\psi_1 + \frac{E_c}{4}\left(|\psi_1|^2-|\psi_2|^2\right)\psi_1 - J\psi_2,
\end{equation}
\begin{equation}
i\hbar\frac{d \psi_2}{dt} = +\frac{\delta}{2}\psi_2 + \frac{E_c}{4}\left(|\psi_2|^2-|\psi_1|^2\right)\psi_2 - J\psi_1.
\end{equation}
Under the transformation, $E_c \rightarrow -E_c$, $J \rightarrow -J$ and $\delta \rightarrow -\delta$, the Hamiltonian (18) $H$ becomes as $-H$. This means that the dimensionless Hamiltonian $H_{d}=H/J$ is invariant under the transformation. In this article, the coupling strength $J$ is assumed to be non-negative.

Using the atomic numbers $n_j$ and the phases $\theta_j$, the complex amplitudes can be expressed in the form of $\psi_j(t)=\sqrt{n_j}\exp\left(i\theta_j\right)$. Thus, one can introduce two new variables, the fractional population imbalance
\begin{equation}
z = \frac{n_1-n_2}{n_1+n_2}=\frac{n_1-n_2}{N},
\end{equation}
and the relative phase
\begin{equation}
\theta = \theta_2-\theta_1.
\end{equation}
From Eqs. (19) and (20), one can find $(\theta, z)$ obeying
\begin{equation}
\frac{d \theta}{dt}=\frac{E_cNz}{2}+\frac{2Jz}{\sqrt{1-z^2}}\cos\theta -\delta,
\end{equation}
\begin{equation}
\frac{d z}{dt}=-2J\sqrt{1-z^2}\sin\theta,
\end{equation}
where we have assumed the Planck constant $\hbar = 1$. Clearly, the equations of motion for $(\theta, z)$ are classical Hamiltonian equations for
\begin{equation}
H_{p}\left(\theta,z\right) = \frac{E_c N}{4}  z^2 + \delta z - 2J \sqrt{1-z^2}\cos \theta,
\end{equation}
which describes a non-rigid pendulum. Here, $(\theta, z)$ are a pair of canonical coordinates corresponding to the angular momentum and the angular variable for a classical pendulum. Thus, the unified MF Hamiltonian for BJJs is equivalent to a classical non-rigid pendulum Hamiltonian.

Although Eqs. (19) and (20) are equivalent to Eqs. (23) and (24), numerical divergence appears in Eq. (23) if $z=\pm 1$ [i.e. all atoms stay in only one of the two modes]. Therefore, for systems involving $z=\pm 1$, one has to choose Eqs. (19) and (20) for numerical simulation.

\subsubsection{Rabi oscillation, macroscopic quantum self-trapping and plasma oscillation}

\noindent The stationary states of BJJs described by Hamiltonian (18) correspond to the fixed points for the classical non-rigid pendulum described by Hamiltonian (25). Mathematically, the fixed points for $(\theta, z)$ are determined by $\frac{d\theta}{dt} =0$ and $\frac{d z}{d t} = 0$, and their general solutions can be written in forms of Jacobian and Weierstrassian elliptic functions~\cite{Smerzi-PRL-1997,Raghavan-PRA-1999}.

Similar to the electronic current in a SJJ, one can define the atomic current in a BJJ as,
\begin{equation}
I=\frac{d \left(n_2-n_1\right)}{d t} = N \frac{d z}{d t} = -2NJ\sqrt{1-z^2}\sin\theta.
\end{equation}
For the BJJ described by the Hamiltonian (18), due to the conservation of its total number of atoms in two BECs, there is no phenomenon of a direct current between two BECs. Although its atomic currents are alternating currents, its phase evolution is not as same as a SJJ with AC Josephson effects, in which the relative phase varies linearly with time. Firstly, in addition to the running-phase modes induced the asymmetry $\delta$ corresponding to the bias voltage in a SJJ, as mentioned above, running-phase modes can also appear in a symmetric BJJ with strong nonlinearity. Secondly, a BJJ  may support Josephson effects of AC atomic currents and oscillating relative phase. By loading atomic BECs into an asymmetric double-well potential and a washboard potential, the AC and DC Josephson effects in atomic BECs have been experimentally demonstrated~\cite{Levy-Nature-2007}.

For a given system of parameters $N$, $E_c$, $J$ and $\delta$, from Eqs. (23) and (24), its stationary states $(\theta^*, z^*)$ can be obtained by solving
\begin{equation}
\frac{E_cNz}{2}+\frac{2Jz}{\sqrt{1-z^2}}\cos\theta -\delta =0,
\end{equation}
\begin{equation}
-2J\sqrt{1-z^2}\sin\theta=0.
\end{equation}
Obviously, from Eq. (28), all stationary states request the relative phase $\theta=k\pi$ with integers $k$. The states of even integers $k=2j$ and odd integers $k=(2j+1)$ are called as in-phase and $\pi$-states, respectively. For simplicity, we will focus on discussing the region of $-\pi \leq \theta \leq +\pi$.

(1) \textsl{Rabi oscillation.} -- For a symmetric system without nonlinearity, $\delta=0$ and $E_c=0$, there are two stationary states $\left(\theta^*=0, z^*=0\right)$ and $\left(\theta^*=\pi, z^*=0\right)$, which have no population imbalance. Its general solutions are periodic oscillations around one of two stationary states with a frequency $\omega_R = 2J$. Actually, this linear system is similar to the Rabi model for a two-level system coupled by classical lasers~\cite{Gerry-Book-2005}. Therefore, these periodic oscillations around a stationary state are called as Rabi oscillations. In Fig. 5, we show the Rabi oscillations in a linear Bose-Josephson system described the Hamiltonian (18) of $J=1$, $\delta=0$, $E_c=0$ and $N=1000$. In which, the black solid dots denote the stationary states.
\vspace{2mm}
\begin{figure}[htp]
\center
\includegraphics[width=1.0\columnwidth]{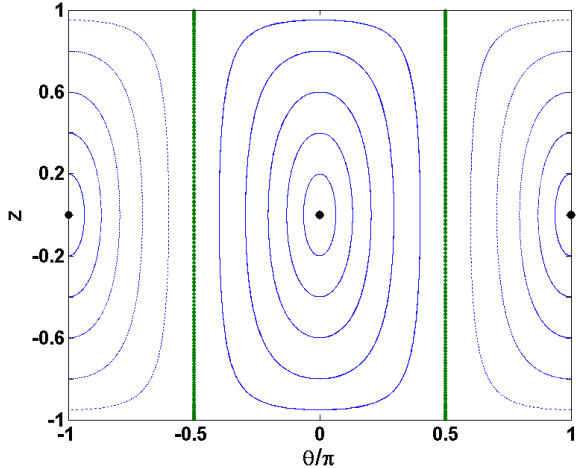}\caption{Rabi oscillations in a linear Bose-Josephson system described by the Hamiltonian (18) of $J=1$, $\delta=0$, $E_c=0$ and $N=1000$.}
\end{figure}

(2) \textsl{MQST.} -- For a symmetric system with non-zero $E_c$, in addition to the two normal stationary states with no population imbalance, two new stationary states appear if $E_c^2N^2 > 16J^2$. If $E_c>0$, two new stationary states are $\pi$-phase states, $\left(\theta^*=\pi, z^*=\pm \sqrt{1-\frac{16J^2}{E_c^2N^2}}\right)$. If $E_c<0$, two new stationary states are in-phase states, $\left(\theta^*=0, z^*=\pm \sqrt{1-\frac{16J^2}{E_c^2N^2}}\right)$. The non-zero population imbalance is a direct signature of MQST~\cite{Kenkre-PRB-1986,Smerzi-PRL-1997,Raghavan-PRA-1999,Kuang-PRA-2000,Hu-PRA-2000}. The critical point, where the two new stationary states appear in addition to the normal stationary state, is as same as the bifurcation point for a Hopf bifurcation~\cite{Lee-PRA-2004}. Usually, the general solutions are periodic oscillations enclosing at least a stationary state. However, due to the nonlinearity, there may exist a new type of solution, running-phase mode, in which the population oscillates periodically and the relative phase changes monotonously.

In Fig. 6, we show the Rabi oscillations and MQST in a nonlinear Bose-Josephson system described by the Hamiltonian (18) of $J=1$, $\delta=0$, $E_c=0.01$ and $N=1000$. The stable and unstable stationary states are denoted by black and red solid dots, respectively. In addition to the normal stationary states of no population imbalance, there appear nontrivial stationary states of non-zero population imbalance, which are self-trapped states. Correspondingly, there are three different types of non-stationary states: (i) Rabi oscillations in the green orbit of $\langle z \rangle_t =0$ and $\langle d\theta/dt \rangle_t =0$; (ii) running-phase MQST between the green and red orbits of $\langle z \rangle_t \neq 0$ and $\langle d\theta/dt \rangle_t \neq 0$, which is labeled as MQST-A ; and (iii) MQST in the red orbits of $\langle z \rangle_t \neq 0$ and $\langle d\theta/dt \rangle_t =0$, which is labeled as MQST-B. Here, $\langle A \rangle_t$ denotes the time averaged value for the variable $A$.

\vspace{2mm}
\begin{figure}[htp]
\center
\includegraphics[width=1.0\columnwidth]{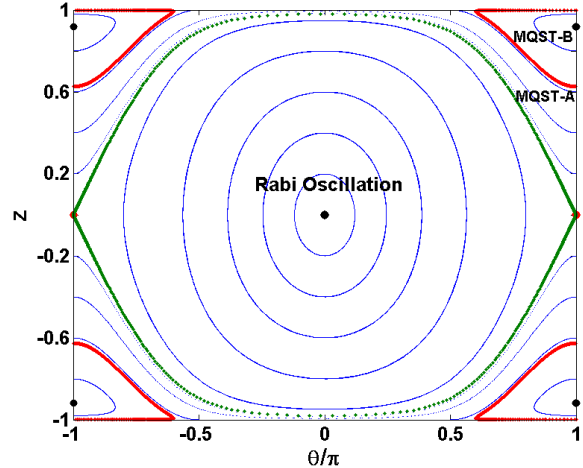}
\caption{Rabi oscillations and MQST in a nonlinear Bose-Josephson system described by the Hamiltonian (18) of $J=1$, $\delta=0$, $E_c=0.01$ and $N=1000$.}
\end{figure}

(3) \textsl{Plasma oscillation.} -- Around a stationary state, a BJJ may have small harmonic oscillations corresponding to plasma oscillations in a SJJ~\cite{Smerzi-PRL-1997,Raghavan-PRA-1999,Barone-Book-1982}. The solution around a stationary state $(\theta^*,z^*)$ can be written in form of
\begin{equation}
\theta=\theta^*+\hat{\theta}, ~~~~~~~~~z=z^*+\hat{z},
\end{equation}
 where $\hat{z}$ and $\hat{\theta}$ are small oscillations around $(\theta^*,z^*)$. Inserting this solution into Eqs. (23) and (24) and keeping only the linear terms of $\hat{z}$ and $\hat{\theta}$, one will obtain the following linearized equations,
\begin{equation}
\frac{d \hat{\theta}}{d t} =\frac{E_cN}{2}\hat{z} +\frac{2J\cos{\theta^*}}{\left[1-\left(z^*\right)^2\right]^{3/2}}\hat{z} -\frac{2J\sin{\theta^*}}{\sqrt{1-\left(z^*\right)^2}}\hat{\theta},
\end{equation}
\begin{equation}
\frac{d \hat{z}}{d t} = \frac{2Jz^*\sin{\theta^*}}{\sqrt{1-\left(z^*\right)^2}}\hat{z} -\left[2J\sqrt{1-\left(z^*\right)^2}\cos{\theta^*}\right]\hat{\theta}.
\end{equation}
As all stationary states satisfy $\sin{\theta^*}=0$, the terms of $\sin{\theta^*}$ disappear in the linearized equations.

For a linear system, due to $E_c=0$ and $z^*=0$, the linearized equations read as
\begin{equation}
\frac{d \hat{\theta}}{d t} = \left[2J\cos{\theta^*}\right]\hat{z}, ~~~~~~\frac{d \hat{z}}{d t} = -\left[2J\cos{\theta^*}\right]\hat{\theta}.
\end{equation}
Clearly, these small oscillations are Rabi-like oscillations of an angular frequency $\omega_{p} = 2J$, which is just the Rabi frequency.

For a nonlinear system around the normal stationary states of $z^*=0$, the linearized equations become as
\begin{equation}
\frac{d \hat{\theta}}{d t} = \left[\frac{E_cN}{2} + 2J\cos{\theta^*}\right]\hat{z}, ~~\frac{d \hat{z}}{d t} = -\left[2J\cos{\theta^*}\right]\hat{\theta},
\end{equation}
which describe the small oscillations of an angular frequency $\omega_{p}=2J \sqrt{1+\frac{E_cN}{4J\cos{\theta^*}}}$. Mathematically, for stationary states satisfying $\frac{E_cN}{4J\cos{\theta^*}}<-1$, the angular frequency $\omega_{p}$ may take imaginary values. The appearance of imaginary $\omega_{p}$ means instability of a stationary state and divergence of all small perturbations to this stationary state. Thus, for an unstable stationary state, there is no small oscillation around it.

For a nonlinear system of $E_c>0$ and $\delta=0$, the in-phase stationary state of $\cos{\theta^*}=+1$ and $z^*=0$ is stable, but the anti-phase stationary state of $\cos{\theta^*}=-1$ and $z^*=0$ becomes unstable if $\left|E_cN\right| > 4J$. For a nonlinear system of $E_c<0$ and $\delta=0$, the anti-phase stationary state of $\cos{\theta^*}=-1$ and $z^*=0$ is stable, but the in-phase stationary state of $\cos{\theta^*}=+1$ and $z^*=0$ becomes unstable if $\left|E_cN\right| > 4J$. Actually, the transition from stable to unstable of a normal stationary state accompanies with the appearance of two additional stable stationary states of non-zero population imbalance. This confirms that the bifurcation from normal to self-trapped states is a Hopf bifurcation~\cite{Lee-PRA-2004}.

For a strongly nonlinear system of $\left|E_cN\right| > 4J$ and $\delta =0$, the small oscillations around the stationary states of non-zero population imbalance $z^*=\pm \sqrt{1-\frac{16J^2}{E_c^2N^2}}$ obey
\begin{equation}
\frac{d \hat{\theta}}{d t} =-\frac{E_cN}{2}\left[\frac{E_c^2 N^2}{16J^2}-1\right]\hat{z},
~~
\frac{d \hat{z}}{d t} = \left[\frac{8J^2}{E_c N}\right]\hat{\theta}.
\end{equation}
These small oscillations are sinusoidal oscillations of an angular frequency $\omega_{p} = 2J \sqrt{\frac{E_c^2 N^2}{16J^2}-1}$.

Different from a linear system, the plasma frequency of a nonlinear system not only depends on the Josephson coupling $J$, but also depends on the nonlinearity $E_c$ and the relative phase $\theta^*$. In a strongly nonlinear system, there are phase-dependent stationary states of self-trapped non-zero population imbalance and small oscillations around these stationary states. Due to the competition between the nonlinearity and the Josephson coupling, the plasma frequency gradually decreases to zero when the system approaches to the critical point, where self-trapping appears.

\subsubsection{Shapiro resonance and chaos}

\noindent In the presence of a periodic driving field, if the energy difference of a particle (or a composite particle such as a Cooper pair) at two sides of the potential barrier can be compensated by the driving field, the system exhibits resonant tunnelling phenomena named as the Shapiro resonance~\cite{Raghavan-PRA-1999,Barone-Book-1982}. In a SJJ subjected to a bias voltage, $V = V_0 + V_1 \cos\left(\omega_d t\right)$, Shapiro resonances occur when the driving frequency satisfies $n\omega_d = \omega_{ac}$. Here, $n$ are positive integers and $\omega_{ac}=2eV_0/\hbar$ is the angular frequency for the AC Josephson oscillations in a SJJ with a DC bias voltage $V=V_0$. The resonance condition indicates that the energy difference is compensated by the energy of $n$ ``photons" from the driving field. This means that the physical pictures for Shapiro resonances and photon-assisted tunnelling~\cite{Eckardt-PRL-2005,Sias-PRL-2008,Xie-PRA-2010} have no any difference.

In a BJJ, the asymmetry $\delta$ acts as the role of the bias voltage $V$ in a SJJ. Similarly, we assume the asymmetry has a constant component and a periodic component, that is, $\delta = \delta_0 + \delta_1 \cos{\omega_d t}$. Such an asymmetry can be realized by controlling double-well asymmetry (or detuning to resonance) for external (or internal) BJJs. Introducing a transformation
\begin{equation}
\varphi_1 \left(t\right) = \psi_1 \left(t\right) \cdot \exp\left[+\frac{i\delta_1}{2} \int  \cos\left(\omega_d t\right) dt \right],
\end{equation}
\begin{equation}
\varphi_2 \left(t\right) = \psi_2 \left(t\right) \cdot \exp\left[-\frac{i\delta_1}{2} \int  \cos\left(\omega_d t\right) dt \right],
\end{equation}
the coupled discrete GPEs (19) and (20) read as
\begin{equation}
i\hbar\frac{d \varphi_1}{dt} = -\frac{\delta_0}{2}\varphi_1 + \frac{E_c}{4}\left(|\varphi_1|^2-|\varphi_2|^2\right)\varphi_1 - \hat{J} \varphi_2,
\end{equation}
\begin{equation}
i\hbar\frac{d \varphi_2}{dt} = +\frac{\delta_0}{2}\varphi_2 + \frac{E_c}{4}\left(|\varphi_2|^2-|\varphi_1|^2\right)\varphi_2 - \hat{J}^*\varphi_1,
\end{equation}
with the new coupling strengthes
\begin{eqnarray}
\hat{J} &=& J \exp\left[+i\delta_1 \int  \cos\left(\omega_d t\right) dt \right]\nonumber\\
 &=& J \exp\left[+i\frac{\delta_1}{\omega_d} \sin\left(\omega_d t\right) \right]\nonumber
\end{eqnarray}
and $\hat{J}^*$ is the complex conjugate for $\hat{J}$. Using the ordinary Bessel functions $B_k\left(x\right)$, the coupling strength $\hat{J}$ can be expanded as,
\begin{equation}
\hat{J} = J \sum_{k=-\infty} ^{+\infty} \textrm{e}^{i k \omega_d t} B_k\left(x\right),
\end{equation}
with $x=\delta_1/\omega_d$. Thus the Hamiltonian (18) becomes
\begin{eqnarray}
&H&= \frac{\delta}{2}\left(n_2-n_1\right) + \frac{E_c}{8}\left(n_2-n_1\right)^2 \nonumber\\
&-&J\sum_{k=-\infty}^{+\infty} B_k\left(x\right)\left(\textrm{e}^{i k \omega_d t} \varphi_1^*\varphi_2+ \textrm{e}^{-i k \omega_d t} \varphi_2^*\varphi_1\right),
\end{eqnarray}
with $n_j=\varphi_j^*\varphi_j=\left|\varphi_j\right|^2$.

For a linear system of zero $E_c$, if coupling term can be regarded as a small perturbation, Shapiro resonances are expected to take place when the energy of an integer number of ``photons" matches the static asymmetry, $n \hbar \omega_d = \delta_0$. Near a n-photon resonance, the detuning $\Delta\left(\omega\right) = n\omega_d - \delta$ is very small. Using the rotating-wave approximation, in the vicinity of a n-photon resonance, the fast oscillating terms with $k \neq n$ in the above Hamiltonian can be ignored. That is, the inter-mode coupling is dominated by the term of $B_n \left(\delta_1/\omega_d\right)$. Therefore, at the n-photon resonance, the driven asymmetric system is approximately equivalent to a undriven symmetric system of an effective coupling strength $J_{\textrm{eff}} = J B_n \left(\delta_1/\omega_d\right)$.

For an actual system under a weak driving, a resonance may occur if a rational multiple of the driving frequency matches the intrinsic frequency of the undriven system. Since the intrinsic frequency depends on both the coupling strength and the nonlinearity strength, the actual resonance positions are shifted away from $n \hbar \omega_d = \delta_0$~\cite{Eckardt-PRL-2005}. Generally, if resonance condition is not satisfied and the driving is sufficiently weak, the oscillations are quasi-periodic ones including two frequency component: the intrinsic oscillation frequency and the driving frequency.

We now consider systems of strong driving fields which can not treated as perturbations. Under a strong driving, the time evolution of a linear system will be quasi-periodic oscillations, which include the intrinsic oscillation frequency and the driving frequency~\cite{Wu-PRL-2007}. For a nonlinear system, chaotic oscillations may appear if the driving is sufficiently strong~\cite{Lee-PRA-2001,Hai-PRE-2002,Abdullaev-PRA-2000}. In Fig. 7, we show Poincare sections for a driven nonlinear Bose-Josephson system (18) with $J=1$, $E_c=0.01$, $N=1000$ and $\delta (t) =\delta_1 \cos\left(2\pi t\right)$. In our simulation, for a given initial condition, we record a data per driving period. In the Poincare sections, a curve or a circle of infinite dots corresponds to a regular motion and a region of infinite dots corresponds to chaotic motions. For a weak driving, the oscillations are regular, see panel (a). When the driving strength $\delta_1$ increases, chaos appear in the vicinity of the unstable stationary state for the undriven system and the region of chaotic motions increases with the driving strength. However, even in the sea of chaos, it is possible to find some islands of regular motions. In our figure, the chaotic sea is denoted by the region of green dots, regular motions in the small regular islands surrounded by chaotic sea are denoted by curves/circles of red dots, and other regular motions are denoted by curves/circles of blue dots.

\vspace{2mm}
\begin{figure}[htp]
\center
\includegraphics[width=1.0\columnwidth]{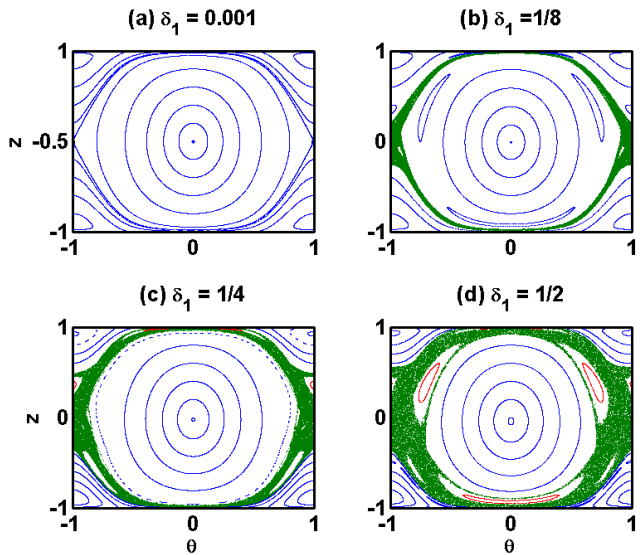}
\caption{Poincare sections for a Bose-Josephson system (18) with $J=1$, $E_c=0.01$, $N=1000$ and time-dependent asymmetry $\delta (t) =\delta_1 \cos\left(2\pi t\right)$. (a) $\delta_1 = 0.001$, (b) $\delta_1 = 1/8$, (c) $\delta_1 = 1/4$ and (d) $\delta_1 = 0.5$.}
\end{figure}

It is also possible to find exact Floquet states for systems driven by some periodic modulations~\cite{Xie-PRA-2007,Luo-PRA-2008,Xie-PRA-2009}. These Floquet states
have no linear counterparts due to their existence requires a nonzero MF interaction~\cite{Xie-PRA-2007}. These nonlinear Floquet states may be stable or unstable. Usually, the stale Floquet states stay around the node points of the undriven system and the unstable Floquet states are enclosed by the chaotic motions around the saddle points of the undriven system. There is a triangular structure in the quasienergy band, which corresponds to the onset of a localization phenomenon called as the coherent destruction of tunneling~\cite{Luo-PRA-2008}.

\subsubsection{Universal mechanism for slow passages through a critical point}

\noindent Mathematically, slow passages through a bifurcation point exhibit memory effects induced by bifurcation delay~\cite{Baer-SIAM-1989,Berglund-JPA-1999}. Physically, slow passages through a critical point of vanished excitation gap connect with the breakdown of adiabaticity and the appearance of non-adiabatic excitations (also called as defect modes)~\cite{Dziarmaga-AdvPhys-2010}. When the system approaches to the critical point, the adiabaticity breaks down and the generated defect modes follows Kibble-Zurek (KZ) mechanism~\cite{Dziarmaga-AdvPhys-2010,Kibble-JPA-1976,Zurek-Nature-1985,Zurek-PhysRep-1996}. As mentioned in Section 2.4.2, for a BJJ described by the Hamiltonian (18), there is a phase transition from normal to self-trapped states, which corresponds to a Hopf bifurcation~\cite{Lee-PRA-2004}. Below, we present the slow passage through the critical point between the normal and self-trapped states~\cite{Lee-PRL-2009}.

For simplicity, we assume the Josephson coupling strength $J \geq 0$ and only discuss the MF ground-states for symmetric BJJs in this section. For a symmetric BJJ of positive $E_c$, the normal-to-MQST transition occurs in anti-phase states, which are not its MF ground-states. For a symmetric BJJ of negative $E_c$, the normal-to-MQST transition occurs in in-phase states, which are its MF ground-states. To push the system across the critical point, one may adjust the nonlinearity $E_c$ via Feshbach resonances~\cite{Feshbach-AnnPhys-1958,Fano-PhysRev-1961,Chin-RMP-2010} or vary the Josephson coupling strength $J$. For external BJJs, the Josephson coupling strength $J$ can be controlled via tuning the potential barrier~\cite{Shin-PRL-2004,Schumm-NatPhys-2005,Schumm-QIP-2006,Albiez-PRL-2005,Gati-JPB-2007}. For internal BJJs, the Josephson coupling strength $J$ can be controlled via tuning the intensity of the coupling lasers~\cite{Hall-PRL-1998,Smerzi-EPJB-2003}.

To explore the normal-to-MQST transition in ground-states, one has to select a system of $E_c<0$, whose critical point is determined by $E_cN = - 4 J$. For this system across the critical point, its ground state changes from $(\psi_2^{GS},\psi_1^{GS})=(\sqrt{L},\sqrt{L})\exp(i\theta_j)$ to $(\sqrt{L-l_s},\sqrt{L+l_s})\exp(i\theta_j)$ with $L=N/2$ and $l_s=\pm\sqrt{L^2-4J^2/E_c^2}$. Clearly, the normal-to-MQST transition is a type of spontaneous symmetry breaking related to internal degrees of freedom~\cite{Lee-PRA-2004,Lee-PRL-2009,Ueda-AIPConfProc-2006}.

In a slow passage through the critical point, there are two characteristic times: the \emph{reaction time} and the \emph{transition time}. The \emph{reaction time}, $\tau_r(t)=\hbar/\triangle_g(t)$, describes how fast the system follows eigenstates of its instantaneous Hamiltonian. The \emph{transition time}, $\tau_t(t)=\triangle_g(t)/|d\triangle_g(t)/dt|$, depicts how fast the system is driven. The adiabatic condition keeps valid if $\tau_r(t)<\tau_t(t)$. Otherwise, if $\tau_r(t)>\tau_t(t)$, the adiabatic condition becomes invalid and defect modes appear.

The Bogoliubov theory is a useful method for analyzing the collective excitations over the ground states~\cite{Griffin-Book-1995,Pethick-Book-2002,Pitaevskii-Book-2003,Ueda-Book-2010,Ozeri-RMP-2005}. In a BJJ described by Hamiltonian (18), due to its spatial degrees of freedom have been frozen, the excitations will not change the spatial distribution but only change the population distribution. Therefore, under the MF description, the perturbed ground state read as
\begin{equation}
\psi_{j}(t)= \left[\chi_{j}+\delta\psi_{j}(t)\right]\exp\left(-\frac{i}{\hbar}\mu t\right)
\end{equation}
with $\psi_j^{GS} = \chi_{j} \exp\left(-\frac{i}{\hbar}\mu t\right)$ and the ground state chemical potential $\mu$. Inserting the above perturbed ground state into coupled GPEs (19) and (20), one will obtain the linearized equations for the perturbations,
\begin{equation}
i\hbar\frac{\partial}{\partial t}\left|\delta\psi(t)\right\rangle =[q]\left|\delta\psi(t)\right\rangle +[p]\left|\delta\psi^{*}(t)\right\rangle,
\end{equation}
with Dirac kets
\begin{equation}
\left|\delta\psi(t)\right\rangle =\left(\begin{array}{c}
\delta\psi_{1}(t)\\
\delta\psi_{2}(t)\end{array}\right),~~\left|\delta\psi^{*}(t)\right\rangle =\left(\begin{array}{c}
\delta\psi_{1}^{*}(t)\\
\delta\psi_{2}^{*}(t)\end{array}\right).
\end{equation}
The two coefficient matrices
\begin{equation}
\left[q\right] =\left(\begin{array}{cc}
q_{11} & q_{12}\\
q_{21} & q_{22}
\end{array}\right),
~~
\left[p\right] =\left(\begin{array}{cc}
p_{11} & p_{12}\\
p_{21} & p_{22}
\end{array}\right)
\end{equation}
have elements
$$q_{11}=-\mu-\frac{\delta}{2}+\frac{E_c}{4}\left(2\left|\chi_{1}\right|^2 -\left|\chi_{2}\right|^2\right),$$
$$q_{12}= -J -\frac{E_c}{4}\chi_{2}^{*}\chi_{1},~~q_{21}= -J -\frac{E_c}{4}\chi_{1}^{*}\chi_{2},$$
$$q_{22}= -\mu +\frac{\delta}{2}+\frac{E_c}{4}\left(2\left|\chi_{2}\right|^2 -\left|\chi_{1}\right|^2\right),$$
$$p_{11} = \frac{E_c}{4}\chi_{1}^{2},~~p_{12} = p_{21} = -\frac{E_c}{4}\chi_{2}\chi_{1},~~p_{22} = \frac{E_c}{4}\chi_{2}^{2}.$$
As the complex amplitudes $\chi_{j}$ for the ground state are determined by the parameters, given the formula above, all matrix elements are determined by the parameters.

The perturbations $\delta\psi_{j}(t)$ can be written in form of
\begin{equation}
\delta\psi_{j}(t)= \sum_{l} \left[u_{jl}\exp(-i\omega_l t)+v^{*}_{jl}\exp\left(+i\omega_l t\right)\right]
\end{equation}
with the excitation frequencies $\omega_l$, the complex amplitudes $(u_{jl}, v_{jl})$ and the complex conjugate $v^{*}_{jl}$ for $v_{jl}$. Substituting the expansion (45) into Eq. (42), one can find
\begin{equation}
\hbar\omega_l \left|u\right\rangle _{l} = [q]\left|u\right\rangle _{l}+[p]\left|v\right\rangle _{l},
\end{equation}
by comparing the coefficients for terms of $\exp(-i\omega_l t)$. Similarly, comparing the coefficients for terms of $\exp(+i\omega_l t)$, one will get
\begin{equation}
-\hbar\omega_l \left|v^{*}\right\rangle _{l}=[q]\left|v^{*}\right\rangle _{l}+[p]\left|u^{*}\right\rangle _{l}.
\end{equation}
Here, the Dirac kets are defined as
\begin{equation}
\left|\beta\right\rangle _{l}=\left(\begin{array}{c}
\beta_{1l}\\
\beta_{2l}\end{array}\right),
~~\left|\beta^{*}\right\rangle _{l}=\left(\begin{array}{c}
\beta_{1l}^{*}\\
\beta_{2l}^{*}\end{array}\right)
\end{equation}
for $\beta=u~\textrm{and}~v$. Combine the above two equations together, one will obtain the eigen-equation for the excitation modes,
\begin{equation}
\hbar\omega_{l}\left(\begin{array}{c}
\left|u\right\rangle_{l}\\
\left|v\right\rangle_{l}\end{array}\right)=\left(\begin{array}{cc}
[q] & [p]\\
-[p*] & -[q*]\end{array}\right)\left(\begin{array}{c}
\left|u\right\rangle_{l}\\
\left|v\right\rangle_{l}\end{array}\right).
\end{equation}
In which, $[A^{*}]$ is the conjugate matrix of $[A]$.

Solving the eigen-equation for a symmetric system who has $\delta=0$, one will find that, apart from a trivial gapless excitation mode, there exists a gapped excitation mode over the ground state. The energy gap for the gapped excitation reads as
\begin{equation}
\Delta_{g}=\hbar \omega_g = \left\{
\begin{array}{cc}2J \sqrt{1+\frac{E_cN}{4J}},&~\textrm{for}~J \geq J_c,\\
2J \sqrt{\frac{E_c^2 N^2}{16J^2}-1}, &~\textrm{for}~J \leq J_c,
\end{array}
\right.
\end{equation}
where $J_c=\left|E_c N/4\right| = -E_c N/4$ is the critical Josephson coupling strength. Clearly, the excitation gap $\Delta_g$ gradually decreases to zero when $J$ approaches to $J_c$. Additionally, it is easy to find that the excitation frequency $\omega_g$ for the gapped mode is just the Plasma oscillation frequency $\omega_p$ discussed in Section 2.4.2. This is because that both $\omega_g$ and $\omega_p$ are obtained from the linearized equations for the perturbations.

We now discuss the dynamical mechanism near the critical point $J=J_c$. We assume $J(t)=J_c(1\pm t/\tau_{q})=J_c\pm\alpha t$ is a linear function of time $t$, where $\tau_q$ is the \emph{quenching time}. The relative coupling, $\varepsilon=|[J(t)-J_c]/J_c|=|t|/\tau_q$, corresponds to the relative temperature in KZ theory~\cite{Dziarmaga-AdvPhys-2010,Kibble-JPA-1976,Zurek-Nature-1985,Zurek-PhysRep-1996}. The vanished gap at the critical point means the divergence of the \emph{reaction time}. In a slow passage through the critical point, the adiabatic evolution becomes invalid if $\tau_r(t) > \tau_t(t)$~\cite{Dziarmaga-AdvPhys-2010,Kibble-JPA-1976,Zurek-Nature-1985,Zurek-PhysRep-1996}. The slow passage divides into three regions: one non-adiabatic region sandwiched by two adiabatic ones, see Fig. 8. The two boundaries are determined by $\tau_r(t) = \tau_t(t)$.

In the normal region, $J(t)>J_c$, introducing $\hat{J}(t)=J(t)-J_c$, we have the \emph{reaction time},
\begin{equation}
\tau_r=\frac{1}{2\sqrt{\hat{J}(\hat{J}+J_c)}}=\frac{1}{\sqrt{\varepsilon(1+\varepsilon)}}\tau_0,
\end{equation}
and the \emph{transition time},
\begin{equation}
\tau_t=\frac{2\hat{J}(\hat{J}+J_c)}{J_c(2\hat{J}+J_c)}\tau_q = \frac{2\varepsilon(1+\varepsilon)}{1+2\varepsilon}\tau_q,
\end{equation}
where $\tau_0=1/\left(2 J_c\right)$. By solving $\tau_r=\tau_t$, one can find the relation,
\begin{equation}
\tau_q=\frac{2\varepsilon+1}{2[\varepsilon(\varepsilon+1)]^{3/2}}\tau_0.
\end{equation}
For a slow passage, $\tau_q\gg 1$, using the Taylor formulation, one will find
\begin{equation}
\varepsilon\approx 2^{-2/3}\tau_0^{2/3}\tau_q^{-2/3}, \   \ |t|\approx 2^{-2/3}\tau_0^{2/3}\tau_q^{1/3}.
\end{equation}

Similarly, in the MQST region, $J(t)<J_c$, introducing $\hat{J}(t)=J_c-J(t)$, we have the \emph{reaction time},
\begin{equation}
\tau_r=\frac{1}{2\sqrt{\hat{J}(2J_c-\hat{J})}}=\frac{1}{\sqrt{\varepsilon(2-\varepsilon)}}\tau_0,
\end{equation}
and the \emph{transition time},
\begin{equation}
\tau_t=\frac{\hat{J}(2J_c-\hat{J})}{J_c(J_c-\hat{J})}\tau_q =\frac{\varepsilon(2-\varepsilon)}{1-\varepsilon}\tau_q.
\end{equation}
Solving $\tau_r=\tau_t$, we have the relation
\begin{equation}
\tau_q=\frac{1-\varepsilon}{[\varepsilon(2-\varepsilon)]^{3/2}}\tau_0.
\end{equation}
For a slow passage, $\tau_q \gg 1$, it is easy to find
\begin{equation}
\varepsilon\approx 2^{-1}\tau_0^{2/3}\tau_q^{-2/3}, \   \ |t|\approx 2^{-1}\tau_0^{2/3}\tau_q^{1/3}.
\end{equation}

Based upon the above analysis on slow passages through the critical point, the universal scalings $\varepsilon\sim\tau_0^{2/3}\tau_q^{-2/3}$ and $|t|\sim\tau_0^{2/3}\tau_q^{1/3}$
recover the KZ mechanism~[6,8]: $\varepsilon\sim\tau_0^{1/(1+z\nu)}\tau_q^{-1/(1+z\nu)}$
and $|t|\sim\tau_0^{1/(1+z\nu)}\tau_q^{z\nu/(1+z\nu)}$ with $z=1$ and $\nu=1/2$, for the
continuous quantum phase transitions~\cite{Dziarmaga-AdvPhys-2010, Zurek-PRL-2005, Dziarmaga-PRL-2005, Polkovnikov-NatPhys-2008}. This means that the normal-to-MQST transition is a continuous quantum phase transition.

In Fig. 8, we show our numerical results for a Bose-Josephson system (18) with $E_c=-0.01$, $N=1000$ and $\delta =0$. The initial state is chosen as a MQST ground state for $J=1.5$ and then $J$ is linearly swept through the critical point of $J=J_c=2.5$. The reaction time $\tau_r$ only relies on the parameters and diverges at the critical point. But the transition time $\tau_t$ depends on both the parameters and the quenching time $\tau_q$. For a larger $\tau_q$, a smaller non-adiabatic region will be found. Due to the divergence of $\tau_r$ induced by the gapless excitation at the critical point, the adiabaticity is always broken down when the system approaches to the critical point. The defect excitations appear after the adiabatic condition becomes invalid. The oscillation amplitudes decrease with the increase of $\tau_q$. That is, the fast passage processes of shorter quenching times have large oscillation amplitudes.

\vspace{2mm}
\begin{figure}[htp]
\center
\includegraphics[width=1.0\columnwidth]{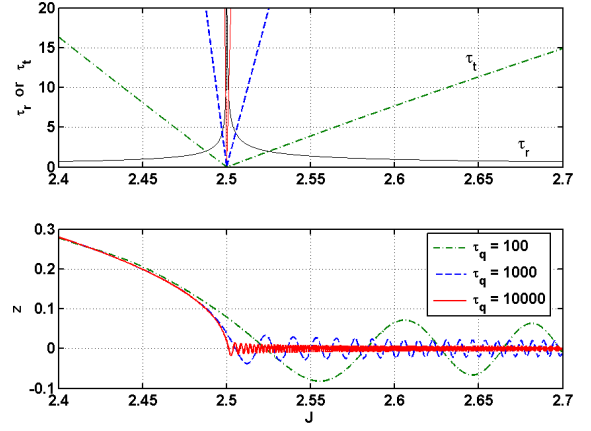}
\caption{Dynamics of slow passages through the critical point for the normal-to-MQST transition in a Bose-Josephson system (18) with $E_c=-0.01$, $N=1000$ and $\delta =0$. The above panel shows how the reaction time $\tau_r$ and the transition time $\tau_t$ depend on the Josephson coupling strength $J$ and the quenching time $\tau_q$. The bottom panel shows the non-adiabatic dynamics of the fractional population imbalance $z$.}
\end{figure}

In addition to the above slow passage through the critical point induced by varying the ratio between the Josephson coupling and the nonlinear interaction, there are several interesting works on adiabatic evolutions and non-adiabatic excitations in nonlinear LZ tunneling processes induced by varying the asymmetry ~\cite{Wu-PRA-2000,Liu-PRA-2002,Jona-Lasinio-PRL-2003,Liu-PRL-2003,Wu-PRL-2006}. For a system of strong nonlinearity, due to the appearance of the loop structure~\cite{Wu-PRA-2000,Liu-PRA-2002}, the adiabatic evolution breaks down~\cite{Liu-PRL-2003} and the semiclassical and adiabatic limits become incommutable~\cite{Wu-PRL-2006}. Qualitatively, the adiabaticity of such a nonlinear evolution requires to vary slowly the parameters with respect to the Bogoliubov excitation frequencies~\cite{Liu-PRL-2003}.

Besides the studies on adiabatic dynamics in BJJs, the adiabatic dynamics in atom-molecular BECs has also been studied~\cite{Pu-PRL-2007,Ling-PRA-2007}. In the absence of inter-particle collisions, the adiabatic condition has been derived with a linearization procedure, in which the collective modes obey the usual orthonormalization relation~\cite{Pu-PRL-2007}. In contrast to the linear quantum systems, if the nonlinearity from two-body collisions can not be ignored, one has to consider the Bogoliubov collective excitation modes~\cite{Ling-PRA-2007}.

\section{Many-body quantum interferometry}

\noindent Beyond the MF theory, the many-body quantum effects become important in systems of low dimensionality, systems possessing a strict symmetry constraint, degeneracy, or strong interaction~\cite{Lewenstein-AdvPhys-2007,Bloch-RMP-2008}. Examples include low dimensional gases, condensates of spinor atoms, rapidly rotating condensates, condensates with strong atom-atom interaction, and ultracold atoms confined in double-well potentials or optical lattices.

To explore the many-body quantum effects, one can employ a second quantization theory. That is, the systems are treated as an ensemble of discrete quantized particles rather than continuous fields. Within the second quantization formalism, the systems of multiple coupled modes obey Hubbard-like Hamiltonians~\cite{Jaksch-PRL-1998,Duan-PRL-2003,Lee-PRL-2004,Wu-JOSAB-2006}, and several many-body quantum phenomena including quantum squeezing~\cite{Orzel-Science-2001,Esteve-Nature-2008}, quantum phase transition~\cite{Greiner-Nature-2002a}, and dynamics and collapse and revival~\cite{Greiner-Nature-2002b} etc have been observed via manipulating Bose atoms in optical lattices. In particular, several many-body states of two-mode systems have been used to implement high-precision quantum interferometry~\cite{Sorensen-Nature-2001, Lee-PRL-2006, Boixo-PRL-2008, Pezze-PRL-2009, Gross-Nature-2010, Riedel-Nature-2010, Choi-PRA-2008a, Choi-PRA-2008b, Boixo-PRA-2009, Tacla-PRA-2010, Boixo-AIPConProc-2009}. Moreover, it has also explored several fundamental noises in quantum interferometry with one-dimensional Bose atomic gases~\cite{Polkovnikov-PNAS-2006, Imambekov-FermiSchool-2007, Hofferberth-Nature-2007, Hofferberth-NaturePhys-2008}.

In this section, we focus on the many-body quantum interferometry via two-mode systems of Bose condensed atoms, which could be described by the unified model for BJJs. Firstly, we give the many-body model in second quantization and discuss some key characteristics of its ground states. Then, we show how to generate spin squeezing and then use it for quantum interferometry. Lastly, we discuss how to prepare NOON states and then use them for quantum interferometry.

\subsection{BJJ in second quantization}

\noindent In the second quantization theory, replacing the complex amplitudes $\psi_j$ ($\psi_j^*$) with the bosonic operators  $\hat{b}_j$ ($\hat{b}_j^+$), the MF Hamiltonian (18) becomes a two-mode Bose-Hubbard model,
\begin{eqnarray}
H &=& \frac{\delta}{2}\left(\hat{n}_2-\hat{n}_1\right) +\frac{E_{c}}{8}\left(\hat{n}_2-\hat{n}_1\right)^{2}\nonumber\\
&& -J\left(\hat{b}_{2}^{\dag}\hat{b}_{1}+\hat{b}_{1}^{\dag}\hat{b}_{2}\right),
\label{pythag}
\end{eqnarray}
with the number operators $\hat{n}_j = \hat{b}_{j}^{\dag}\hat{b}_{j}$ and the total number of atoms $N = \langle\hat{N}\rangle = \langle\hat{n}_1 + \hat{n}_2\rangle$. In the Fock basis $\{ \left|n_1, n_2\right\rangle \}$, any arbitrary many-body quantum states of the above Hamiltonian can be written in form of $\left|\Psi\right\rangle = \sum_{n_1,n_2}^{} C_{n_1,n_2}\left|n_1, n_2\right\rangle$ with $C_{n_1,n_2}$ denoting the probability amplitudes for $\left|n_1, n_2\right\rangle$. Obviously, this Hamiltonian conserves the total number of atoms $N$ due to $\left[\hat{N}, H\right] = 0$.

The many-body ground states of the quantized BJJ described by Hamiltonian (59) depend on its parameters and the total number of atoms $N$. If the Hamiltonian is dominated by the Josephson coupling term of $J$, the ground state is a SU(2) coherent state. If it is dominated by the nonlinear term of positive $E_c$, the ground state is a single Fock state $\left|\frac{N}{2}, \frac{N}{2}\right\rangle$ for even $N$ or a twin Fock state $\frac{1}{\sqrt{2}} \left(\left|\frac{N-1}{2}, \frac{N+1}{2}\right\rangle + \left|\frac{N+1}{2}, \frac{N-1}{2}\right\rangle\right)$ for odd $N$. If it is dominated by the nonlinear term of negative $E_c$, there are two degenerated ground states $\left|N, 0\right\rangle$ and $\left|0, N\right\rangle$, so that any superposition of these two states including the NOON state $\frac{1}{\sqrt{2}}\left(\left|N, 0\right\rangle + \left|0, N\right\rangle\right)$ is also a ground state.

A quantized BJJ of $N$ atoms can be treated as a spin-$\frac{N}{2}$ particle in an external magnetic field. By introducing the spin operators,
\begin{eqnarray}
\hat{S}_{x}&=\frac{1}{2}(\hat{b}_{2}^{\dag}\hat{b}_{1}+\hat{b}_{1}^{\dag}\hat{b}_{2}),\nonumber\\
\hat{S}_{y}&=\frac{1}{2i}(\hat{b}_{2}^{\dag}\hat{b}_{1}-\hat{b}_{1}^{\dag}\hat{b}_{2}),\nonumber\\
\hat{S}_{z}&=\frac{1}{2}(\hat{b}_{2}^{\dag}\hat{b}_{2}-\hat{b}_{1}^{\dag}\hat{b}_{1}),
\end{eqnarray}
the Hamiltonian (59) is equivalent to a giant spin model~\cite{Pu-PRL-2002,Lee-PRA-2003,Wu-PRA-2003},
\begin{equation}
H = -B_{x}\hat{S}_{x} - B_{z}\hat{S}_{z} + D\hat{S}_{z}^{2},
\label{pythag}
\end{equation}
with the transverse magnetic field $B_{x} = 2J$, the longitudinal magnetic field $B_{z} = -\delta$ and the axial anisotropy energy $D = \frac{E_{c}}{2}$.

\vspace{2mm}
\begin{figure}[htp]
\center
\includegraphics[width=1.0\columnwidth]{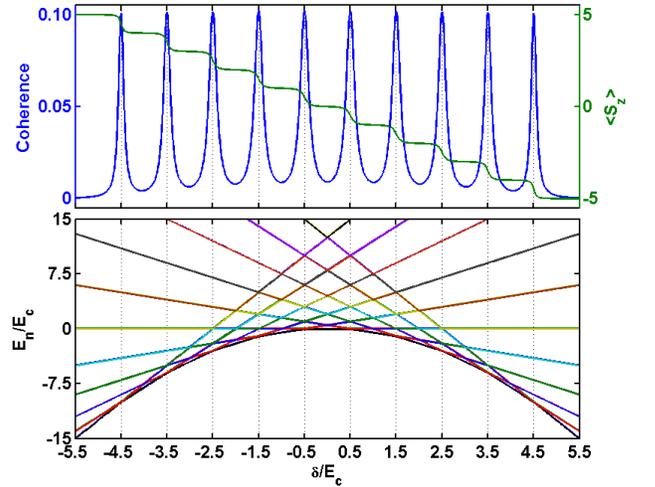}
\caption{Interaction blockade, resonant tunneling and energy spectrum for a quantized BJJ of $N=10$ and $E_c/J = 10$. The first row shows the coherence and the expectation value $\langle S_z \rangle$ versus the asymmetry strength $\delta/E_c$. The second row shows how $E_n/E_c$ depends on $\delta/E_c$ with $E_n$ denoting the eigenenergy for the n-th eigenstate.}
\end{figure}

In a quantized BJJ of weak Josephson coupling, the competition between asymmetry and nonlinear interaction will result interaction blockade and resonant tunnelling~\cite{Lee-EPL-2008,Capelle-PRL-2007,Dounas-Frazer-PRL-2007,Cheinet-PRL-2008}. The single-atom resonant tunneling occurs when the asymmetry is balanced by the interaction energy, that is,
\begin{equation}
\frac{\delta_{RT}}{2}n + \frac{E_c}{8}n^2=  \frac{\delta_{RT}}{2}n' + \frac{E_c}{8}\left(n'\right)^2,
\end{equation}
with $n=n_2-n_1$, $n'=n_2'-n_1'$, $n_2'=n_2\pm1$ and $n_1'=n_1\mp1$. Simplifying the balance condition, one can easily obtain
\begin{equation}
\delta_{RT} + E_c \left(m+\frac{1}{2}\right) =0,
\end{equation}
with $m=\frac{n}{2}=\{-\frac{N}{2},-\frac{N}{2}+1,\cdots, \frac{N}{2}-1\}$ and $N=n_1+n_2$. This means that there are $N$ resonant peaks for the single-particle tunneling in a system of $N$ atoms. Actually, due to the positivity of $E_c$, the resonant peaks for the single-particle tunneling correspond to the avoided crossings between the ground and first-excited states. Between two resonant peaks, the quantum tunneling between two modes is blockaded by the large gap originated from the interaction energy $E_c$. This is a kind of interaction blockade similar to the Coulomb blockade of electrons in quantum dots. The phenomena of resonant tunneling and interaction blockade have been observed in experiments~\cite{Cheinet-PRL-2008}.

\subsection{Spin squeezing and quantum interferometry}

\noindent Quantum squeezing~\cite{Walls-Book-2008,Scully-Book-1997} is a consequence of Heisenberg's uncertainty relations. A state is squeezed when the fluctuations of one variable is reduced below the symmetric limit at the expense of the increased fluctuations of the conjugate variable. The quantum squeezed states of photons have provided important applications in high-precision interferometry~\cite{Caves-PRD-1981,Bondurant-PRD-1984} and have stimulated extensive interests in both theoretical and experimental quantum optics.

To illustrate the conception of squeezing, we consider a harmonic oscillator described by the Hamiltonian,
\begin{equation}
H=\frac{\hat{p}^{2}}{2m}+\frac{1}{2}m\omega^{2}\hat{x}^{2}.
\end{equation}
In units of $\hbar=1$, one can introduce the creation operator
\begin{equation}
\hat{a}^{\dag}\equiv\sqrt{\frac{m\omega}{2}}\hat{x}-i\sqrt{\frac{1}{2m\omega}}\hat{p},
\end{equation}
and the annihilation operator
\begin{equation}
\hat{a}\equiv\sqrt{\frac{m\omega}{2}}\hat{x}+i\sqrt{\frac{1}{2m\omega}}\hat{p}.
\end{equation}
These two operators obey a bosonic commutation relation $[\hat{a},\hat{a}^{\dag}]=1$. It is convenient to introduce two new dimensionless operators,
\begin{equation}
\hat{X}=\sqrt{2m\omega}\hat{x}=\hat{a}+\hat{a}^{\dag},
\end{equation}
\begin{equation}
\hat{P}=\sqrt{\frac{2}{m\omega}}\hat{p}=\frac{\hat{a}-\hat{a}^{\dag}}{i}.
\end{equation}
Therefore, we have the commutation relation $[\hat{X},\hat{P}]=2i$ and the uncertainty relation $\Delta \hat{X}\Delta \hat{P}\geq 1$. The state of $\Delta \hat{X}=\Delta \hat{P}=1$ is a coherent state~\cite{Zhang-RMP-1990}. Otherwise, the state is a squeezed one if $\Delta \hat{X}<1$ or $\Delta \hat{P}<1$.

For systems represented by spin or angular momentum operators, such as the Hamiltonian (61), the commutation relations for $\hat{S}_{x,y,z}$ read as
\begin{equation}
[\hat{S}_{i},\hat{S}_{j}]=i\epsilon_{ijk}\hat{S}_{k},
\label{pythag}
\end{equation}
where $\epsilon_{ijk}$ is the Levi-Civita symbol and $\left(i, j, k\right)$ are indices for three Cartesian components along directions $\left(\vec{e}_x, \vec{e}_y, \vec{e}_z\right)$. The commutation relations leads to Heisenberg's uncertainty relations
\begin{equation}
\Delta \hat{S}_{i} \cdot \Delta \hat{S}_{j} \geq \frac{1}{2}\left|\langle \hat{S}_{k}\rangle \right|.
\end{equation}
In analogy to a harmonic oscillator, a coherent spin state (CSS)~\cite{Ma-arXiv-2011,Zhang-RMP-1990} is a collective state describing a collection of spins that are perfectly aligned or polarized in one direction $\vec{e}_k$, in which the fluctuations along two other orthogonal directions $\Delta \hat{S}_{i} = \Delta \hat{S}_{j} = \frac{1}{\sqrt{2}}\left|\langle \hat{S}_{k}\rangle\right|^{\frac{1}{2}}$. Otherwise, a spin state is squeezed if $\Delta \hat{S}_{i}$ or $\Delta \hat{S}_{j}$ is smaller than $\frac{1}{\sqrt{2}}\left|\langle \hat{S}_{k}\rangle\right|^{\frac{1}{2}}$~\cite{Ma-arXiv-2011, Zhang-RMP-1990, Kitagawa-PRA-1993}. In language of quantum phase states, a spin-$\frac{N}{2}$ CSS is a phase state for a two-mode Bose system of N particles~\cite{Anglin-PRA-2001}. For a double-well BEC system, the differences between MF and many-body quantum dynamics have been explored by using CSS's~\cite{Weiss-PRL-2008}.

Generally, a spin-$\frac{N}{2}$ system can be regarded as a collective system of N spin-$\frac{1}{2}$ particles. The large spin operators can be constructed by the collective spin operators
\begin{equation}
\hat{S}_{\alpha}=\frac{1}{2}\sum\limits_{l=1}^{N}\hat{\sigma}^{\alpha}_{l},
\end{equation}
where $\alpha=(x,y,z)$ and $\hat{\sigma}^{\alpha}_{l}$ are the Pauli matrices for the $l$-th particle. Thus, the CSS can be expressed as a direct product of N single-particle states
\begin{equation}
\left|\theta,\phi\right\rangle = \bigotimes \limits_{l=1}^{N} \left[\cos\frac{\theta}{2} \left|\downarrow\right\rangle_{l} + e^{i\phi}\sin\frac{\theta}{2} \left|\uparrow\right\rangle _{l}\right],
\end{equation}
where $\left|\downarrow\right\rangle_{l}$ and $\left|\uparrow\right\rangle_{l}$ are spin-down and spin-up states of $\hat{\sigma}^{z}_{l}$ respectively. The means that all N particles point to the same direction $(\theta,\phi)$, which is called as the mean spin direction(MSD). The variance of the components perpendicular to MSD is simply the sum of the variances of $N$ individual particles, and it is $\frac{N}{4}$ since there are no quantum correlations among particles. Spin squeezed states (SSS) are associated with the quantum correlations or entanglement among particles. It is possible to reduce the fluctuations in one direction at the expense of increasing fluctuations in the other direction.

There are several parameters for characterizing the spin squeezing~\cite{Ma-arXiv-2011}. Below, we give the definitions for three squeezing parameters, which are mostly used in related fields.

\begin{itemize}
\item The first squeezing parameter is defined according to the Heisenberg's uncertainty relation~\cite{Walls-PRL-1981}. It is given as
\begin{equation}
\xi_{H}^{2}=\frac{2(\Delta \hat{S}_\alpha)^{2}}{|\langle \hat{S}_{\gamma}\rangle|},\alpha\neq\gamma.
\end{equation}
With this definition, the state is squeezed if $\xi_{H}^{2}<1$.

\item The second squeezing parameter is determined the minimum fluctuation along a particular direction~\cite{Kitagawa-PRA-1993}. It is defined as
\begin{equation}
\xi_{S}^{2}=\frac{\textrm{min}\left(\Delta \hat{S}_{\vec{n}_{\bot}}\right)^{2}}{j/2} =\frac{4\cdot \textrm{min}\left(\Delta \hat{S}_{\vec{n}_{\bot}}\right)^{2}}{N}
\end{equation}
where the spin length $j=N/2$, the unitary vector $\vec{n}_{\bot}$ denotes a direction perpendicular to the MSD and the minimization is over all possible directions $\vec{n}_{\bot}$. This parameter is set to find out the most squeezed direction perpendicular to the MSD. If $\xi_{S}^{2}<1$, the state is squeezed.

\item The third squeezing parameter is the ratio of phase fluctuations for the considered state and a reference CSS~\cite{Wineland-PRA-1992,Wineland-PRA-1994}. It is expressed as
\begin{equation}
\xi_{R}^{2}=\frac{\Delta\phi}{(\Delta\phi)_{CSS}}=\frac{N(\Delta \hat{S}_{\vec{n}_{\bot}})^{2}}{|\langle\hat{S}\rangle|^{2}}.
\end{equation}
Similarly, the state is squeezed if $\xi_{R}^{2}<1$.
\end{itemize}

Now, we discuss how to enhance measurement precision by using spin squeezing. For an example, we consider a Ramsey interferometry of two $\pi/2$-pulses sandwiching a free evolution. The Ramsey interferometry is equivalent to a Mach-Zehnder (MZ) interferometry, in which the two $\pi/2$-pulses act as the two beam splitters. Assuming an unknown phase $\phi$ about to the z-axis is accumulated in the free evolution, the state of the Ramsey interferometer evolves according to
\begin{equation}
\left|\Psi(t)\right\rangle = \hat{U} \left|\Psi(0)\right\rangle,
\end{equation}
with the propagation operator
\begin{eqnarray}
\hat{U}  &=& \exp \left(-i\frac{\pi}{2}\hat{S}_x\right) \exp \left(-i\phi\hat{S}_z\right) \exp \left(-i\frac{\pi}{2}\hat{S}_x\right) \nonumber\\
 &=& \exp \left(-i\phi\hat{S}_y\right) \exp \left(-i\pi\hat{S}_x\right).
\end{eqnarray}

To estimate the phase $\phi$, one may measure a observable $\hat{Q}$, which has $\phi$-dependent expectation values and variances. According to the error propagation formula, the variance of $\phi$ is given by
\begin{equation}
\Delta \phi =  \frac{\left(\Delta \hat{Q}\right)}{\left| \partial \left\langle \hat{Q}\right\rangle /\partial \phi\right|}.
\end{equation}
In the Ramsey interferometry, it is usually to measure $\hat{S}_z$ of the final state. The expectation value of $\hat{S}_z$ is
\begin{eqnarray}
\left\langle \hat{S}_z\right\rangle_{f} = &\langle \Psi(t)|\hat{S}_z|\Psi(t)\rangle\nonumber\\
= &\langle \Psi(0)|\hat{U}_{+} \hat{S}_z \hat{U}|\Psi(0)\rangle\nonumber\\
= & \sin \phi \cdot \langle \hat{S}_x \rangle_{t=0} - \cos \phi \cdot \langle \hat{S}_z \rangle_{t=0}
\end{eqnarray}
and the corresponding variance is
\begin{eqnarray}
\left(\Delta \hat{S}_z \right)^{2}_{f} = &\sin^2 \phi \cdot \left( \Delta \hat{S}_x \right)^{2}_{t=0} + \cos^2 \phi \cdot \left( \Delta \hat{S}_z \right)^{2}_{t=0}\nonumber\\
 &- \sin \left(2\phi\right) \cdot \textrm{Cov}\left(\hat{S}_x,\hat{S}_z\right)_{t=0}
\end{eqnarray}
with
\begin{equation}
\textrm{Cov}\left(\hat{S}_x,\hat{S}_z\right)= \frac{1}{2}\left\langle \hat{S}_x \hat{S}_z + \hat{S}_z \hat{S}_x \right\rangle.
\end{equation}
Thus, we can estimate the unknown phase with the uncertainty
\begin{equation}
\Delta \phi = \frac{\left(\Delta \hat{S}_{z}\right)_{f}} {\left| \partial \left\langle \hat{S}_{z}\right\rangle_{f} /\partial \phi\right|}.
\end{equation}
If the input state is an CSS, the phase sensitivity is
\begin{equation}
(\Delta \phi)_{\textrm{CSS}}=\frac{1}{\sqrt{N}},
\end{equation}
which is the standard quantum limit (SQL)~\cite{Giovannetti-Science-2004}. If the input state is a SSS, the phase sensitivity reads as
\begin{equation}
(\Delta \phi)_{\textrm{SSS}} = \frac{\xi_{R}}{\sqrt{N}}.
\end{equation}
This means that the phase sensitivity could be improved beyond the SQL by utilizing spin squeezing.

Usually, one can generate spin squeezing by nonlinear inter-particle interactions. In 1993, Kitagawa and Ueda demonstrated that the squeezed states of $\xi_{R} \varpropto 1/N^{-1/3}$ (or $1/N^{1/2}$) can be generated by the one-axis (or two-axis) twisting Hamiltonians~\cite{Kitagawa-PRA-1993}. Recently, it has also demonstrated that planar spin squeezed states can be prepared via attractive BECs~\cite{He-PRA-2011}. Therefore, using these squeezed states, the phase sensitivity can beat the SQL provided by the unentangled states. For a two-component atomic BEC, it has proposed that the nonlinear twisting can be accomplished by the s-wave scattering between atoms and the spin squeezing is a useful criterion for many-particle entanglement~\cite{Sorensen-Nature-2001}. Recently, two experimental groups have prepared SSS's of Bose condensed atoms by using one-axis twisting and demonstrated high-precision quantum interferometry beyond the SQL by utilizing these squeezed states~\cite{Gross-Nature-2010, Riedel-Nature-2010}. In a practical BEC system, particle loss~\cite{Li-PRL-2008} and non-condensed atoms have significant effects on the optimal spin squeezing achievable from the inter-atom interactions~\cite{Jin-PRA-2010a,Jin-PRA-2010b,Tikhonenkov-PRA-2010,Maussang-PRL-2010,Sinatra-PRL-2011, Genoni-PRL-2011}.

Additionally, other controlled interactions such as laser-induced interactions~\cite{Hald-PRL-1999,Molmer-PRL-1999,Liu-PRL-2011}, continuous quantum non-demolition measurements~\cite{Kuzmich-PRL-2000} and Coulomb interactions~\cite{Meyer-PRL-2001,Leibfried-Science-2004,Roos-Nature-2006} are also able to generate non-classical entangled states for implementing high-precision interferometry.

\subsection{NOON states and quantum interferometry}

\noindent A NOON state~\cite{Sanders-PRA-1989,Boto-PRL-2000,Lee-JMO-2002} is a many-body quantum state,
\begin{equation}
\left|\textrm{NOON}\right\rangle = \frac{1}{\sqrt{2}}\left(\left|N\right\rangle_{a}\left|0\right\rangle_{b} + \textrm{e}^{i\theta} \left|0\right\rangle_{a}\left|N\right\rangle_{b}\right),
\end{equation}
which represents a superposition of N particles in mode a with zero particles in mode b, and vice versa. If the two modes are regarded as two possible paths for the particles, the NOON state is also called as the path-entangled state. Due to only two modes are involved, the NOON state is equivalent to the N-particle GHZ state,
\begin{equation}
\left|\textrm{GHZ}\right\rangle = \frac{1}{\sqrt{2}}\left(\left|\frac{N}{2},+\frac{N}{2}\right\rangle + \textrm{e}^{i\theta} \left|\frac{N}{2},-\frac{N}{2}\right\rangle\right),
\end{equation}
which is a maximally entangled state.

Because of their entanglement characteristics, NOON and GHZ states can be used to implement Heisenberg-limited interferometry~\cite{Giovannetti-Science-2004}. For an example, we consider a N-particle Ramsey interferometer. If we input a GHZ state,
\begin{equation}
\left|\textrm{GHZ}\right\rangle_{0} = \frac{1}{\sqrt{2}} \left(\left|\frac{N}{2},+\frac{N}{2}\right\rangle_{y} + \left|\frac{N}{2},-\frac{N}{2}\right\rangle_{y}\right),
\end{equation}
after the Ramsey process, the final output state is
\begin{equation}
\left|\textrm{GHZ}\right\rangle_{f} = \frac{1}{\sqrt{2}} \left(\left|\frac{N}{2},+\frac{N}{2}\right\rangle_{y} + \textrm{e}^{-iN\phi} \left|\frac{N}{2},-\frac{N}{2}\right\rangle_{y}\right).
\end{equation}
For this final state, because the initial state has zero expectations for $\hat{S}_x$ and $\hat{S}_z$, it is not possible to extract any information about $\phi$ by measuring $\hat{S}_z$. To estimate the phase $\phi$, one may measure the parity operator $\hat{P} = \prod _{l=1}^{N} \sigma_{l}^{z}$, where $\sigma_{l}^{z}$ is the z-component Pauli matrix for the l-th particle~\cite{Bollinger-PRA-1996}. With some analytical calculations, one can easily find that the final expectation values and variances of $\hat{P}$ after the Ramsey procedure are given as
\begin{equation}
\left\langle \hat{P} \right\rangle_{f} = \cos \left(N\phi\right), \left(\Delta \hat{P} \right)^2_f = \sin^{2}\left(N\phi\right).
\end{equation}
Therefore, the phase sensitivity is determined by
\begin{equation}
\Delta \phi =  \frac{\left(\Delta \hat{P}\right)_{f}}{\left| \partial \left\langle \hat{P}\right\rangle_{f} /\partial \phi\right|} = \frac{1}{N},
\end{equation}
which reaches the Heisenberg limit $1/N$.

To estimate $\phi$ from the final state (88), one can also measure the operator $\hat{A} = \left|\Uparrow\rangle \langle \Downarrow \right| + \left|\Downarrow\rangle \langle \Uparrow \right|$ with $\left|\Uparrow\right\rangle = \left|\frac{N}{2},+\frac{N}{2}\right\rangle_{y}$ and $\left|\Downarrow\right\rangle = \left|\frac{N}{2},-\frac{N}{2}\right\rangle_{y}$. This operator has similar expectation values and variances for the parity operator $\hat{P}$.

Although, theoretically, NOON states may enhance the measurement precision to the Heisenberg limit. But the measurement of the parity $\hat{P}$ is very difficult for systems of large $N$ because it requires the ability to distinguish between odd and even numbers of spin-down particles. Combining ideas from precision spectroscopy and quantum information processing, it has demonstrated the Heisenberg-limited interferometry in an experiment with three beryllium ions~\cite{Leibfried-Science-2004}, in which the nonlinear interaction $\chi \hat{S}_x^2$ from the laser-induced two-ion transitions has been used to prepare the NOON state and transfer the phase information into amplitude information. However, these protocols seems very difficult to scale to many-body systems of large $N$.

For the systems of Bose condensed atoms, there are several methods for preparing high NOON states. Similar to prepare the spin squeezing, the high NOON states can be periodically generated by in the time evolutions~\cite{Yorke-PRL-1986, Castin-LectureNotes-2001, Yang-CTP-2009}. For double-well BEC systems, the macroscopic superposition states of all atoms in one well (a type of NOON states) could be generated by phase engineering~\cite{Mahmud-JPB-2003, Mahmud-PRA-2005,Leung-arxiv-2010}. The efficient creation of NOON states has also been demonstrated by utilizing quantum tunnelling~\cite{Watanabe-PRA-2007, Watanabe-PRA-2010}. Moreover, it has also suggested that the high NOON states can be generated by an effective interaction between two atoms from coherent Raman processes through a two-atom molecular intermediate state~\cite{Helmerson-PRL-2001}. Recently, it finds that the tunneling resonances in a titled double-well BEC system lead to a dynamical scheme for creating few-body NOON-like macroscopic superposition states~\cite{Carr-EPL-2010}. Below, we focus on discussing how to prepare high NOON states with adiabatic evolutions and then use them to accomplish a MZ interferometry~\cite{Lee-PRL-2006}.

Based upon the quantized BJJ, which is described by the giant-spin Hamiltonian (61) with $D<0$ (i.e. $E_c<0$), one of us suggest an alternative protocol for preparing high NOON states and then performing the Heisenberg-limited MZ interferometry via adiabatic evolutions through critical points~\cite{Lee-PRL-2006}. In principle, this protocol is suitable for many-body systems of thousands of Bose condensed atoms. In analogy to an optical MZ interferometer, the two collective spin states $\left|\frac{N}{2},-\frac{N}{2}\right\rangle$ and $\left|\frac{N}{2},+\frac{N}{2}\right\rangle$ are used as the two paths, and the beam splitters are realized by the adiabatic processes through the bifurcation point, see Fig. 10. Below, we give some key points of this protocol.

\vspace{3mm}
\begin{figure}[htp]
\center
\includegraphics[width=1.0\columnwidth]{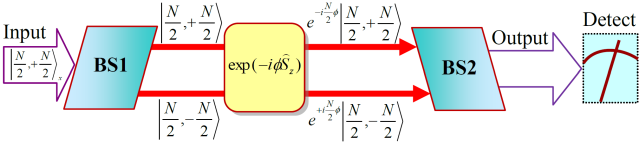}
\caption{Schematic diagram for the adiabatic MZ interferometry via a quantized BJJ described by the giant-spin Hamiltonian (61). The input state is a SU(2) coherent spin state along the x-axis. The beam splitters BS1 and BS2 are achieved by the adiabatic processes through the critical point for a bifurcation between normal and self-trapped states.}
\end{figure}

The initial state is a CSS along the x-axis, which is the ground state for the giant-spin Hamiltonian of $B_z=0$, $B_x>0$ and $D=0$. The desired NOON state is an superposition of two degenerated ground states for the case of $B_z=0$, $B_x=0$ and $D<0$. It has demonstrated that a Hopf bifurcation from normal to self-trapped states takes place when the ratio $\left|D/B_x\right|$ increases from below to above the critical value $4/N$. This bifurcation, which relates to spontaneous symmetry breaking, can be looked as a continuous quantum phase transition. Keeping $B_z=0$ and $D<0$ unchanged, starting from the CSS along the x-axis, the NOON state can be adiabatically prepared by decreasing $B_x$ from $B_x\gg \left|D\right|$ to zero, which accompanies with the occurrence of a dynamical bifurcation. This is to say, the first beam splitter could be achieved by the adiabatic evolution from $\left|B_x/D\right|\gg 1$ to $\left|B_x/D\right|= 0$. Even in the presence of $B_z$ originated from the asymmetry $\delta$, the NOON state can be still prepared with high fidelity if $\delta < \delta_c$.

After accomplishing the first beam splitter, the Josephson coupling between two modes is completely switched off, the state vector processes about the z-axis in the free evolution described by the unitary operator $\exp\left[-i\omega_0 t \hat{S}_z\right]$ and accumulates a phase $\phi = \omega_0 t$. Here, $\omega_0$ is the transition frequency between states $\left|\downarrow\right\rangle$ and $\left|\uparrow\right\rangle$. Thus the state at the end of the free evolution is
\begin{eqnarray}
\left|\Psi\right\rangle &= \exp\left(-i\phi\hat{S}_z\right)\cdot \frac{1}{\sqrt{2}}\left(\left|\frac{N}{2},-\frac{N}{2}\right\rangle +\left|\frac{N}{2},+\frac{N}{2}\right\rangle\right)\nonumber\\
&=\frac{1}{\sqrt{2}}\left(e^{+i\frac{N}{2}\phi}\left|\frac{N}{2},-\frac{N}{2}\right\rangle +e^{-i\frac{N}{2}\phi}\left|\frac{N}{2},+\frac{N}{2}\right\rangle\right).
\end{eqnarray}

The second beam splitter is achieved by the inverse process of the first beam splitter. That is, keeping $B_z=0$ and $D<0$ unchanged, $B_x$ is adiabatically increased from zero to $B_x\gg \left|D\right|$. After completing the second beam splitter, due to the dynamical symmetry breaking in the beam splitter, the final state is a superposition of the ground and first-excited states for the strong coupling limit $B_x\gg \left|D\right|$ and the phase information of the state (91) are transferred into the amplitude information of the two lowest eigenstates, that is
\begin{equation}
\left|\Psi\right\rangle_{f} = \cos\left(\frac{N}{2}\phi\right)\left|\textrm{GS}\right\rangle +e^{i\theta}\sin\left(\frac{N}{2}\phi\right)\left|\textrm{FES}\right\rangle.
\end{equation}
Here, $\left|\textrm{GS}\right\rangle$ and $\left|\textrm{FES}\right\rangle$ denote the ground and first-excited states, respectively. In the strong coupling limit, $\left|B_x/D\right| \gg 1$, the ground state $\left|\textrm{GS}\right\rangle = \left|\frac{N}{2},+\frac{N}{2}\right\rangle$ is a state of all spins along the positive x-axis direction, and the first-excited state $\left|\textrm{FES}\right\rangle = \left|\frac{N}{2},+\frac{N}{2}-1\right\rangle$ is a state of $(N-1)$ spins along the positive x-axis direction and one spin along the negative x-axis direction.

To estimate the phase $\phi$, we measure the operator
\begin{equation}
\hat{R} = \left|\textrm{FES}\rangle \langle\textrm{FES}\right| - \left|\textrm{GS}\rangle \langle\textrm{GS}\right|.
\end{equation}
This leads to a measured signal and variance given by
\begin{equation}
\left\langle \hat{R} \right\rangle_{f} = - \cos \left(N\phi\right), \left(\Delta \hat{R} \right)^2_f = \sin^{2}\left(N\phi\right).
\end{equation}
Thus, the phase sensitivity is determined by
\begin{equation}
\Delta \phi =  \frac{\left(\Delta \hat{R}\right)_{f}}{\left| \partial \left\langle \hat{R}\right\rangle_{f} /\partial \phi\right|} = \frac{1}{N},
\end{equation}
which reaches the Heisenberg limit $1/N$.

The measurement of $\hat{R}$ requires the ability to distinguish $\left|\textrm{GS}\right\rangle$ and $\left|\textrm{FES}\right\rangle$. However, in the strong coupling limit, it is difficult to directly distinguish these two eigenstates. So that one has to transfer these two states into distinguishable states. Fortunately, one can use the similarity of the two lowest eigenstates for symmetric and asymmetric BJJs in the strong coupling limit and the non-degeneracy of the two lowest eigenstates for an asymmetric BJJ in the strong nonlinear limit. After accomplishing the second beam splitter, the system stays in the strong coupling limit, a proper asymmetry $\delta$ is suddenly applied and then
slowly decrease $B_x$ from $B_x \gg |D|$ to zero. To keep the populations in $\left|\textrm{GS}\right\rangle$ and $\left|\textrm{FES}\right\rangle$ unchanged, the process must avoid the dynamical bifurcation and so that the asymmetry must satisfy $|\delta|<|E_{c}/2|$ (the best choice is $|\delta|=|E_{c}/4|$). Lastly, when $B_x$ decreases to zero, $\left|\textrm{GS}\right\rangle = \left|N/2,-N/2\right\rangle$ and $\left|\textrm{FES}\right\rangle = \left|N/2,+N/2\right\rangle$, which can be easily distinguished.

Comparing with the conventional Ramsey interferometry, in which the beam splitters are achieved by $\pi/2$ pulses, in the interferometry protocol discussed above, the beam splitters are achieved by the adiabatic evolutions through the critical point for a Hopf bifurcation (also the critical point for a quantum phase transition). Moreover, in the adiabatic interferometry protocol, the beam splitters have important functions in addition to splitting and recombination. Besides splitting the input state, the first beam splitter prepares the desired NOON state for the Heisenberg-limited interferometry. Besides recombining the two paths, the second beam splitter transfers the phase information into amplitude information.

\section{Conclusions and perspectives}

\noindent In summary, we have given a review of recent progresses in nonlinear quantum interferometry via Bose condensed atoms. The nonlinearity originates from the atom-atom interaction, which can be tuned by applying external magnetic fields or lasers. Besides various exotic phenomena caused by the atom-atom interaction, the atom-atom interaction can be used to generate multi-particle entangled states, such as spin squeezed states and NOON states, which have been suggested for implementing high-precision measurements beyond the SQL provided by the non-entangled states.

For the matter-wave interferometry based on the MQC of atomic BECs, the system can be well described by the MF theory. In 1D systems, matter-wave solitons may be gradually generated in the interference process and more solitons will be generated for systems of more strong nonlinearity. Similar to the SJJs, the BJJs of linearly coupled BECs are widely used to detect and exploit the MQC of BECs. In addition to the Rabi oscillation, a direct signature of MQC, the MQST and chaos may appear in BJJs of strong nonlinearity. By tuning the effective nonlinearity, the BJJ undergoes a symmetry-breaking transition from normal to self-trapped states, in which the critical point depends on both the effective nonlinearity and relative phase. The slow passage dynamics across the critical point shows universal behaviors obeying the KZ mechanism for a continuous quantum phase transition.

For the quantum interferometry based upon the many-body quantum coherence of Bose condensed atoms, the system must be described by the second quantization theory. Under the second quantization, a BJJ is described by two-mode Bose-Hubbard model, which is equivalent to an anisotropic giant-spin within external magnetic fields. The atom-atom interaction in this system is just the nonlinear term of the one-axis twisting Hamiltonian for generating spin squeezing. So that the atom-atom interaction has been suggested for preparing spin squeezed states and then one can use those states for implementing high-precision interferometry beyond the SQL. Moreover, by using adiabatic processes through the critical point, the NOON state can be prepared and then one can use this maximally entangled state for implementing Heisenberg-limited interferometry. It has suggested an adiabatic MZ interferometry, in which, the beam splitters are realized by the adiabatic processes through the critical point and the two parts of the NOON states are used as two paths.

It has been widely accepted that the multi-particle entanglement can enhance the measurement precision and the inter-particle interaction can be used to generate quantum entanglement. However, to design and build a practical quantum device based upon the quantum interferometry with multi-particle entangled states, there are still several important problems waiting to be solved.

\begin{itemize}
\item Firstly, although only internal hyperfine or external motional states are used for interferometry, for an example an atom clock only use two internal states for interferometry. However, the atomic systems have both internal and external degrees of freedom and they may couple with each other~\cite{Li-EPJB-2009}. Therefore, for interferometry with internal states, it is important to explore the effects from the mechanical motion of the center of mass.

\item Secondly, most of present works assume the systems are closed systems. Actually, almost all practical systems are open systems. Beyond the theoretical frame for closed systems, we have to treat the systems as open systems and explore whether an interferometry protocol is robust against the environment effects. Thus, it is vital to explore the effects from decoherence, temperature~\cite{Sinatra-PRL-2011}, dissipation~\cite{Krauter-PRL-2011, Watanabe-arXiv-2011} and particle losses~\cite{Kacprowicz-NatPhoton-2010} etc.

\item Thirdly, all interferometry protocols involve dynamical processes. It is essential to know whether imperfect excitations appear in the dynamical processes and how the imperfect excitations affect the interferometry.
\end{itemize}

{\bf Acknowledgements:} This work is supported by the NNSFC under Grant No. 11075223, the NBRPC under Grant No. 2012CB821300 (2012CB821305), the NCETPC under Grant No. NCET-10-0850 and the Fundamental Research Funds for Central Universities of China.

\end{document}